\documentclass[a4paper,fleqn,usenatbib]{mnras}

\usepackage{newtxtext,newtxmath}

\usepackage[T1]{fontenc}
\usepackage{ae,aecompl}

\usepackage{graphicx}	
\usepackage{amsmath}	

\title[KT Eri is a Recurrent Nova]{The Nova KT Eri Is a Recurrent Nova With a Recurrence Time-Scale of 40-50 Years}

\author[B. E. Schaefer]{
Bradley E. Schaefer$^{1}$\thanks{E-mail: schaefer@lsu.edu},
Frederick M. Walter$^{2}$, 
Rebekah Hounsell$^{3,4}$, \&
Yael Hillman$^{5}$
\\
$^{1}$Department of Physics and Astronomy, Louisiana State University, Baton Rouge, Louisiana, 70820, USA\\
$^{2}$Department of Physics and Astronomy, Stony Brook University, Stony Brook, New York, 11794, USA\\
$^{3}$University of Maryland, Baltimore County, Baltimore, MD 21250, USA\\
$^{4}$NASA Goddard Space Flight Center, Greenbelt, MD 20771, USA\\
$^{5}$Department of Physics, Ariel University, Ariel, POB 3, 4070000, Israel \\
}

\pubyear{2021}

\begin{document}
\label{firstpage}
\pagerange{\pageref{firstpage}--\pageref{lastpage}}
\maketitle

\begin{abstract}

KT Eridani was a very fast nova in 2009 peaking at $V$=5.42 mag.  We marshal large data sets of photometry to finally work out the nature of KT Eri.  From the {\it TESS} light curve, as confirmed with our radial velocity curve, we find an orbital period of 2.61595 days.  With our 272 spectral energy distributions from simultaneous $BVRIJHK$ measures, the companion star has a temperature of 6200$\pm$500 K.  Our century-long average in quiescence has $V$=14.5.  With the {\it Gaia} distance (5110$^{+920}_{-430}$ parsecs), the absolute magnitude is $M_{V_q}$=$+$0.7$\pm$0.3.  We converted this absolute magnitude (corrected to the disc light alone) to accretion rates, $\dot{M}$, with a full integration of the $\alpha$-disc model.  This $\dot{M}$ is very high at 3.5$\times$10$^{-7}$ M$_{\odot}$/year.  Our search and analysis of archival photographs shows that no eruption occurred from 1928--1954 or after 1969.  With our analysis of the optical light curve, the X-ray light curve, and the radial velocity curve, we derive a white dwarf mass of 1.25$\pm$0.03 M$_{\odot}$.  With the high white dwarf mass and very-high $\dot{M}$, KT Eri must require a short time to accumulate the required mass to trigger the next nova event.  Our detailed calculations give a recurrence time-scale of 12 years with a total range of 5--50 years.  When combined with the archival constraints, we conclude that the recurrence time-scale must be between 40--50 years.  So, KT Eri is certainly a recurrent nova, with the prior eruption remaining undiscovered in a solar gap of coverage from 1959 to 1969.
 
\end{abstract}




\section{Introduction}

KT Eridani (Nova Eridani 2009) appeared as an ordinary nova, discovered by K. Itagaki on 2009 November 25, roughly ten days after the optical maximum.  The entire fast rise, peak, and decline out to the time of discovery was covered photometrically (Hounsell et al. 2010) with the Solar Mass Ejection Imager ({\it SMEI}) instrument onboard the {\it Coriolis} satellite.  KT Eri was a bright nova (peaking near 5.42 mag), very fast (fading from peak by 3.0 mag in 13.6 days), with a `P' (plateau) class light curve, and a He/N spectral class.  The progenitor was quickly identified as a 15th mag star, for which Jurdana-\v{S}epi\'{c} et al. (2012) sought prior nova events on 1012 Harvard plates from 1888 to 1962.  KT Eri was also detected in the radio (O'Brien et al. 2010) and X-ray (Bode 2010; Beardmore et al. 2010; Ness et al. 2015; Sun et al. 2020; Pei et al. 2021).

KT Eri is an interesting and unusual nova in a variety of ways:  (1) Hounsell et al. (2010) measured the first well-sampled light curve of the entire initial fast rise of any nova.  (2) This light curve (see inset of Figure 4 of Hounsell et al. 2010) shows the best example of a pre-maximum halt.  (3) KT Eri shows {\it two} distinct flat plateaus in its light curve, in defiance of all prior examples and outside the formal classification scheme (Strope, Schaefer, \& Henden 2010).  (4) KT Eri has been picked out by various groups as having properties that are hallmarks of being a recurrent nova (RN).

Despite this interest, previously, little was known about the system, and nothing fundamental.  That is, the nature of the companion star, the distance, the binary orbital period, and the nova-to-nova recurrence time-scale were not known with any reliability.  To illustrate, the literature contains identification of the companion star as an A7 {\rm III} giant star, an M {\rm III} red giant, or a low-mass faint red main sequence star.  To further illustrate the previously unknown nature of both the eruption and the companion, it has been claimed to be an ordinary classical nova (CN) {\it inside} the Period Gap by the big catalog of Ritter \& Kolb (2003), an RN with a very long orbital period (e.g., Raj, Banerjee, \& Ashok 2013), while the official VSX catalog lists it as a ``Z And" symbiotic star and a likely eclipsing binary.


\section{Observations}

\subsection{Photometry}

Nova eruptions are defined and recognized by their brightness variations.  Photometry of the system serves to measure many of the critical system properties.  For this purpose, our group has measured a very large data set across in $BVRIJHK$ bands from 1888 to 2020.  Further, we have collected magnitudes from the literature and from various public domain data collections.  A full listing of our photometry sources is presented in Table 1.  The explicit measures from all these sources is given in Table 2, with 9627 lines in the on-line version of the table.

\begin{table*}
	\centering
	\caption{KT Eri photometry data sources}
	\begin{tabular}{lllllll} 
		\hline
		Telescope & Reference & Start Date  &  Start JD   &   End JD   &   Band   &   Number\\
		\hline
Harvard (HCO)	&	This paper	&	1888 Oct 19	&	2410929.8	&	2447564.9	&	B	&	139 (1200$^a$)	\\
Sonneberg	&	This paper	&	1950 Oct 13	&	2433567.5	&	2453380.5	&	B	&	0 (700$^a$)	\\
{\it ASAS}	&	[1]; Pojmanski (1997)	&	2000 Nov 20	&	2451868.7	&	2455166.6	&	V	&	53	\\
Catalina	&	[2]; Drake et al. (2014)	&	2005 Sep 7	&	2453621.2	&	2456580.0	&	CV	&	303	\\
SMEI	&	This paper; Hounsell et al. (2010)	&	2009 Nov 11	&	2455147.5	&	2455162.7	&	CV	&	159	\\
Liverpoool	&	This paper; Hounsell et al. (2010)	&	2009 Nov 13	&	2455149.5	&	2455216.4	&	CV	&	437	\\
AAVSO	&	[3]; [4]	&	2009 Nov 25	&	2455161.0	&	2459277.0	&	UBVRI	&	20, 148, 729, 84, 68 (26194$^b$)	\\
Torun	&	Ilkiewicz et al. (2014)	&	2009 Nov 27	&	2455162.5	&	2455592.4	&	BVRI	&	9, 9, 9, 9	\\
SMARTS	&	This paper, Walter et al. (2012)	&	2009 Dec 3	&	2455168.7	&	2458694.9	&	BVRIJHK	&	936, 639, 635, 639, 431, 429, 461	\\
PTF	&	[5]; Rau et al. (2009)	&	2010 Sep 21	&	2455461.0	&	2456282.7	&	R	&	55	\\
{\it APASS}	&	[6]; [7]; Henden et al. (2012)	&	2011 Jan 29	&	2455590.6	&	2455880.7	&	BVgri	&	4, 4, 4, 4, 4	\\
Cembra	&	Munari \& Dallaporta (2014)	&	2011 Sep 30	&	2455834.7	&	2456373.3	&	BVRI	&	118, 157, 113, 156 	\\
Rozhen	&	Ilkiewicz et al. (2014)	&	2012 Oct 25	&	2456226.4	&	2456226.4	&	BVR	&	1, 1, 1	\\
ZTF	&	[8]; Bellm et al. (2019)	&	2018 Mar 29	&	2458206.6	&	2458881.7	&	gR	&	100, 100	\\
{\it TESS} Sector 5	&	This paper	&	2018 Nov 15	&	2458438.0	&	2458464.1	&	CI	&	1181	\\
Las Cumbres	&	This paper	&	2019 Dec 17	&	2458834.9	&	2458926.0	&	BVRI	&	15, 14, 13, 13	\\
{\it TESS} Sector 32	&	This paper	&	2020 Nov 20	&	2459174.2	&	2459200.1	&	CI	&	1223 (102757$^c$)	\\
		\hline
	\end{tabular}
	
\begin{flushleft}	
$^a$This is the approximate plate count including those  for which dates were not recorded, most of which have KT Eri being invisible below the plate limit \\
$^b$Nightly averages of 26194 magnitudes are tabulated to avoid swamping the data file \\
$^c$The full light curve has 102757 fluxes with 20 second time resolution, with the tabulated light curve binned to near 1800 second resolution \\
References: [1] http://www.astrouw.edu.pl/asas/?page=aasc
[2] http://nunuku.caltech.edu/cgi-bin/getcssconedb\_release\_img.cgi
[3] https://www.aavso.org/data-download
[4] https://www.aavso.org/tags/kt-eri
[5] https://irsa.ipac.caltech.edu/cgi-bin/Gator/nph-scan?mission=irsa\&submit=Select\&projshort=PTF
[6] https://www.aavso.org/aavso-photometric-all-sky-survey-data-release-1
[7] https://www.aavso.org/download-apass-data
[8] https://irsa.ipac.caltech.edu/cgi-bin/Gator/nph-scan?projshort=ZTF
\end{flushleft}

\end{table*}

\begin{table}
	\centering
	\caption{KT Eri photometry (full table with 9627 lines available on-line as supplementary material)}
	\begin{tabular}{lllll} 
		\hline
		HJD & Year & Magnitude  &  Unit   &   Source   \\
		\hline
2413218.5846	&	1895.068	&	15.0	$\pm$	0.15	&	B	&	HCO (A 1268)	\\
2421163.7970	&	1916.821	&	14.5	$\pm$	0.15	&	B	&	HCO (MC 11394)	\\
2421272.5132	&	1917.118	&	14.5	$\pm$	0.15	&	B	&	HCO (MC 12366)	\\
2421522.8347	&	1917.804	&	14.4	$\pm$	0.15	&	B	&	HCO (MC 13859)	\\
2421545.7706	&	1917.867	&	14.7	$\pm$	0.15	&	B	&	HCO (MC 14282)	\\
...	&		&				&		&		\\
2459249.0000	&	2021.094	&	15.3	$\pm$	0.0	&	V	&	AAVSO	\\
2459257.0000	&	2021.116	&	15.3	$\pm$	0.0	&	V	&	AAVSO	\\
2459260.0000	&	2021.124	&	15.4	$\pm$	0.0	&	V	&	AAVSO	\\
2459268.0000	&	2021.146	&	15.2	$\pm$	0.0	&	V	&	AAVSO	\\
2459277.0000	&	2021.170	&	15.0	$\pm$	0.0	&	V	&	AAVSO	\\
		\hline
	\end{tabular}

\end{table}

\subsubsection{Eruption Light Curve}

The eruption light curve is presented in Figure 1.  The fast rise, peak, and initial decline are almost entirely covered by the light curves from the {\it SMEI} satellite (Hounsell et al. 2010).  The {\it SMEI} instrumental response was that of an unfiltered CCD calibrated with V magnitudes of 17 bright stars around the ecliptic (Buffington et al. 2007; Hick, Buffington, \& Jackson 2007).  The start date of the eruption was 2009 November 13.12 UT (JD 2455148.62) at magnitude 8.44$\pm$0.09.  The pre-maximum halt in the fast rise towards maximum occurred at magnitude 6.04$\pm$0.07 starting on 2009 November 13.83 UT (JD 2455149.33$\pm$0.04) and lasted for roughly 0.14 days.  Pre-maximum halts have the theoretical explanation as a ``temporary drop in the energy flux as convection in the expanding, thinning envelope ceases to be efficient near the envelope surface" (Hillman et al. 2014).  The peak was at a magnitude of 5.42$\pm$0.02 on 2009 November 14.67 UT (JD 2455150.17$\pm$0.04).  The peak magnitude and time are placed into one summary table (Table 3), with this serving as a collection place for all the critical properties of KT Eri.

\begin{figure*}
	\includegraphics[width=2.2\columnwidth]{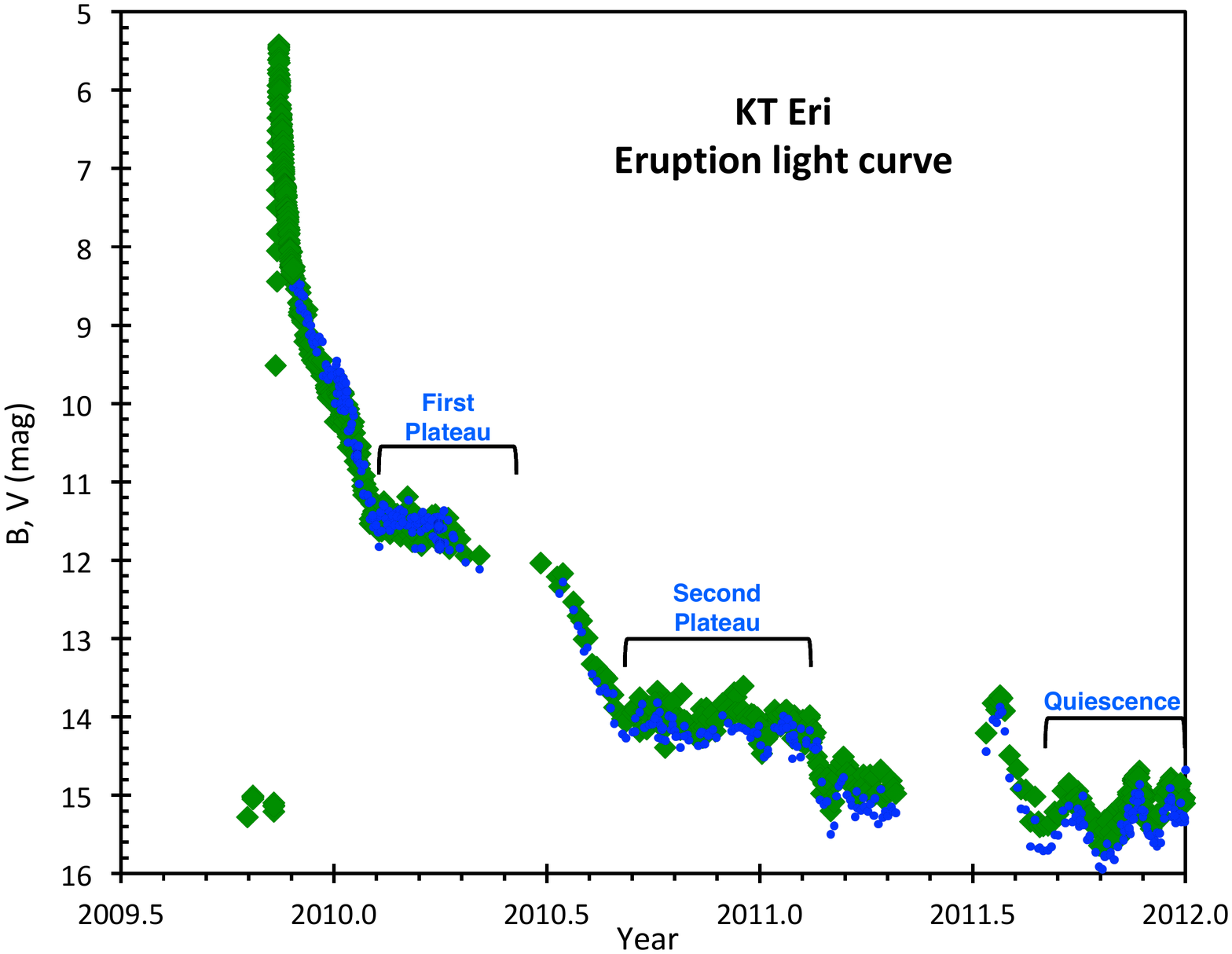}
    \caption{Eruption light curve for KT Eri.  All V-band observations are shown by the green diamonds, and all B-band observations are shown as small blue circles, with the data sources listed in Table 1.  The two color light curves are largely superposed on each other.  The {\it SMEI} light curve shows a nicely-resolved fast rise, a pre-maximum halt, the fast peak, and the rapid initial decline.  (A blow-up figure around the peak is shown in Hounsell et al. 2010.)  Startlingly, {\it two} plateaus are prominent in the light curve, with this being unprecedented.  The star returned to the pre-eruption quiescence level by October 2011.}  
\end{figure*}

KT Eri took 6.6 days after maximum for the V-band light curve to decline by 2 mags, or $t_2$=6.6 days.  The time needed to decline by 3 magnitudes from peak, $t_3$, is 13.6 days for KT Eri.  Specifying time-scales for large magnitude drops will be required for evaluating the historical coverage for prior nova eruptions.  For this, $t_5$ is 61 days, $t_6$ is 80 days, $t_7$=240 days, and $t_8$=277 days.

\begin{table}
	\centering
	\caption{Collected properties of KT Eri)}
	\begin{tabular}{lll} 
		\hline
		Property & Value & Unit   \\
		\hline
Peak magnitude 	&	 5.42$\pm$0.02 	&	 mag 	\\
Peak date 	&	 2455150.17$\pm$0.04 	&	 JD 	\\
Light curve class 	&	 PP, very fast 	&	 - 	\\
Spectral class 	&	 He/N class 	&	 - 	\\
$t_2$ 	&	 6.6$\pm$0.1 	&	 days 	\\
$t_3$ 	&	 13.6$\pm$0.5 	&	 days 	\\
$t_5$ 	&	 61$\pm$3 	&	 days 	\\
$t_6$ 	&	 80$\pm$3 	&	 days 	\\
$t_7$ 	&	 240$\pm$4 	&	 days 	\\
$t_8$ 	&	 277$\pm$4 	&	 days 	\\
$t_9$ 	&	 462$\pm$2	&	 days 	\\
$\langle V_{pre} \rangle$ 	&	 14.5 (13.5 to 15.8) 	&	 mag 	\\
$\langle V_{post} \rangle$ 	&	 15.06 (13.88 to 15.84) 	&	 mag 	\\
$V_{post}$-$V_{pre}$ 	&	 $+$0.56 $\pm$ 0.3	&	mag	\\
Nova amplitude 	&	 9.1$\pm$0.2 	&	 mag 	\\
No eruptions 	&	 1928--1954, after 1969 	&	 year 	\\
$T$ companion star 	&	 6200$\pm$500 	&	 $\degr$K	\\
Radius of companion star 	&	 3.7 $\pm$ 0.3 	&	 R$_{\odot}$	\\
$K$ 	&	 58.4$\pm$7.0 	&	 km s$^{-1}$	\\
$\gamma$ 	&	 -142$\pm$5 	&	 km s$^{-1}$ 	\\
Epoch of conjunction 	&	 2455491.323$\pm$0.053 	&	 HJD 	\\
Orbital period $P$ 	&	 2.61595$\pm$0.00060 	&	 days 	\\
Orbital inclination 	&	 52$\pm$5 	&	 $\degr$	\\
Distance 	&	 5110$^{+920}_{-430}$ 	&	 parsecs 	\\
$E(B-V)$ 	&	 0.082$\pm$0.003 	&	 mag 	\\
$A_V$ 	&	 0.254$\pm$0.008 	&	 mag 	\\
Mass of companion star 	&	 1.0 $\pm$ 0.2 	&	 M$_{\odot}$ 	\\
$M_{WD}$ 	&	 1.25$\pm$0.03 	&	 M$_{\odot}$ 	\\
$M_{V_{peak}}$ 	&	 -8.38$\pm$0.25 	&	 mag 	\\
Century-long $\langle V_q \rangle$ 	&	 14.5$\pm$0.2 	&	 mag 	\\
Range $\langle V_q \rangle$ 	&	 14.0 to 15.3	&	 mag 	\\
Century-long $\langle M_{V_q} \rangle$ 	&	 $+$0.70$\pm$0.3	&	 mag 	\\
Range $\langle M_{V_q} \rangle$ 	&	 $+$0.2 to $+$1.5 	&	 mag 	\\
Century-long $\langle \dot{M} \rangle$ 	&	 3.5$^{+1.8}_{-1.2}$$\times$10$^{-7}$ 	&	 M$_{\odot}$/year 	\\
Range $\langle \dot{M} \rangle$ 	&	 (0.83--8.2)$\times$10$^{-7}$ 	&	 M$_{\odot}$/year 	\\
$\tau_{rec}$ 	&	 40--50 	&	 years 	\\
		\hline
	\end{tabular}

\end{table}

	The eruption light curve shows a prominent plateau, an interval where the light curve suddenly stops its decline for a significant interval.  That is, from days 80--190 after maximum, KT Eri had a nearly constant brightness at $V$$\approx$11.6.  On this basis, we would classify KT Eri as being a `P' class nova light curve (see Strope et al. 2010).  However, KT Eri also has a second prominent plateau from 300--440 days after peak, where KT Eri is constant at a magnitude of $V$=14.1.  A second plateau in any nova light curve was completely unprecedented at the time.{\footnote{This was the first time in which any nova light curve did not exactly fit into the light-curve-classification system.  Even though the number of plateaus required for a `P' classification was not specified, the possibility of a second plateau was not conceived.  Only after KT Eri's eruption did both U Sco and V339 Del display {\it two} plateaus.  A simple solution for the classification system is that any nova with a second plateau is still labelled as `P'.  Alternatively, it might be helpful to distinguish this two-plateau case, in which case we could define a subclass labelled as `PP'.  The important point is that KT Eri displayed an entirely new phenomenon, and such needs an explanation.  Theoretical studies with full physics have produced plateaus, and some of the models produce {\it two} plateaus (Hillman et al. 2014).}}

\subsubsection{Pre-Eruption Light Curve}

The pre-nova system of KT Eri was quickly identified as a $V$$\sim$15 mag star exhibiting 1.8 mag of variability (Drake et al. 2009).  Jurdana-\v{S}epi\'{c} et al. (2012) examined 1012 archival photographic plates dating from 1888 to 1962.  KT Eri was invisible on 517 of these plates, with typical limits of 13th mag.  KT Eri was detected on 495 plates, mostly from 1925 to 1953, and only three detections before 1925.  The native plate sensitivity is close to the Johnson B magnitude system, therefore with comparison stars in the Johnson B-band, the magnitudes returned are closely in the modern B system.  A primary goal was to search for prior eruptions.  (Various workers had pointed to similarities between KT Eri and recurrent novae, and this provided a good motivation to search for eruptions in the prior century.)  No undiscovered nova events were turned up.  Unfortunately, the magnitudes are not quoted in the paper, nor available elsewhere, and details (like a claimed photometric periodicity) cannot be tested with the given light curve.

We have sought pre-eruption images of KT Eri in many archival sources worldwide.  Roughly 1200 useful plates were found at Harvard for the years 1888--1989, many of which were not deep enough to record KT Eri.  Roughly 700 useful plates were found at Sonneberg for the years 1950--2009, with most of the plates yielding non-detections.  The goal was not to measure a light curve, but rather was to quantify the lack of prior eruptions.  Thus, all these plates were looked at individually for KT Eri being bright, but only the dates and limits of plates at the start and end of each observing season were noted.  Almost incidentally, 139 B magnitudes were measured for Harvard plates from 1888--1989.

Our Harvard light curve can be used to test for long term trends before eruptions.  (The Hibernation model predicts systematic pre-eruption brightening, but the time-scale for this is tens-of-thousands of years, see Hillman et al. 2020.  Still, few nova can be tested back as far as KT Eri, then it is worthwhile to look for anything surprising.)  No long term pre-eruption trends are seen.  Another result from this light curve is that the claimed 737 day periodicity (Jurdana-\v{S}epi\'{c} et al. 2012) was not recovered.  This is not surprising because the claimed period is close to exactly-two-years, with this being a common indication of a spurious periodicity related to the data sampling.

\begin{figure}
	\includegraphics[width=1.1\columnwidth]{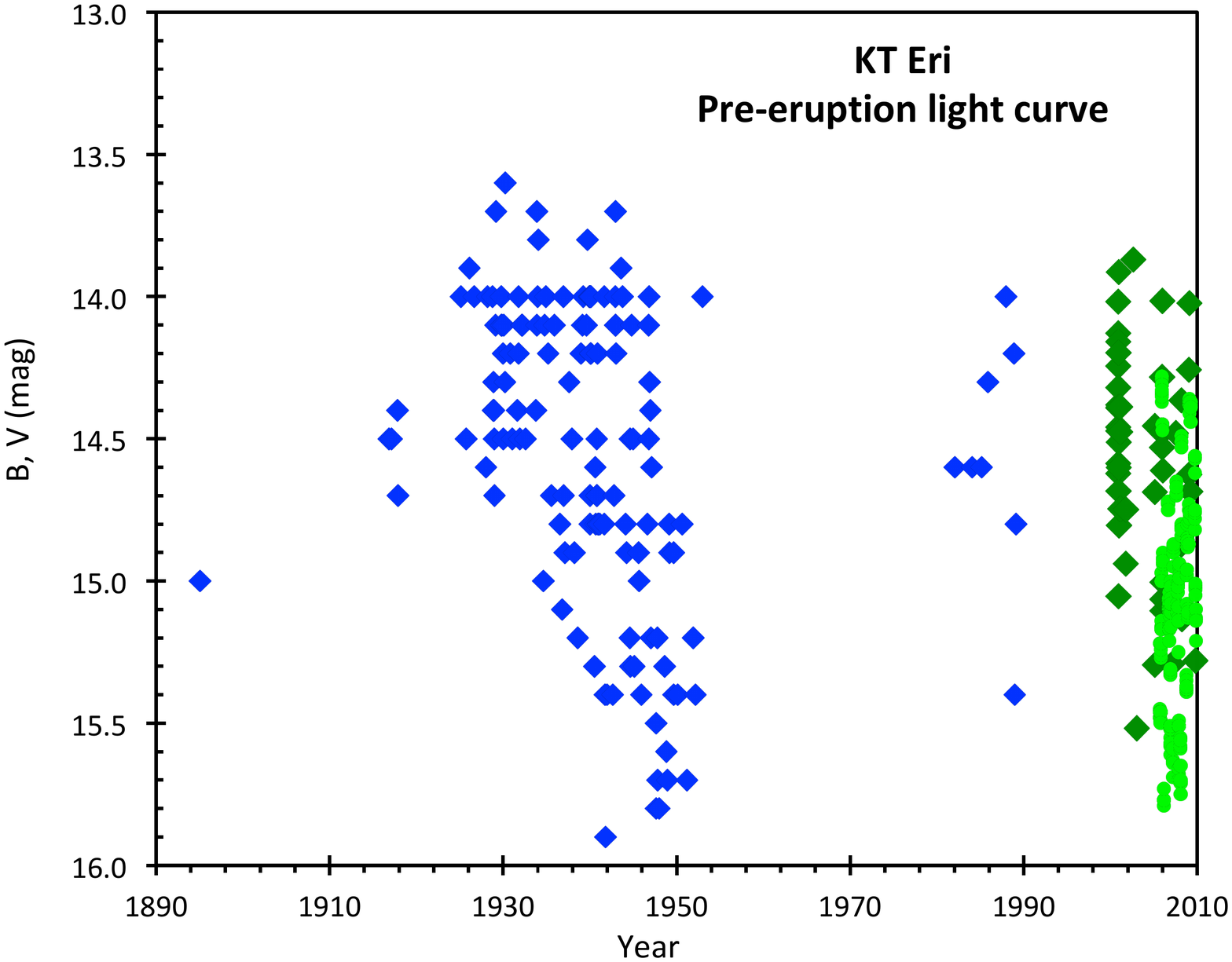}
    \caption{Pre-eruption light curve for KT Eri.  The Harvard $B$-band magnitudes are displayed as blue diamonds, the {\it ASAS} $V$-band magnitudes are in dark green diamonds, and the Catalina $CV$ magnitudes are shown by light green circles.  Importantly, for comparison with the $V$-magnitudes, the $B$-magnitudes need to offset upward by $\langle$B-V$\rangle$=0.23, as taken from the post-eruption data.  One point from this figure is that KT Eri shows no significant long term trends for over a century before its eruption.  Another point is that KT Eri varies greatly around its average of $\langle$V$\rangle$=14.5, from around 13.5 to 15.8 mag in the $V$-system.  Another point is that there are yearly and decadal variations in the quiescent level.}  
\end{figure}

We have collected pre-eruption magnitudes from Harvard, from {\it ASAS}, and the Catalina Sky Survey, see Fig. 2.  KT Eri displays extreme variability, roughly from 13.5 to 15.8 mag in the $V$-band.  Substantial changes are seen in the yearly and decadal averages.  For example, $\langle$B$\rangle$ is 14.19 in the 1920s, 14.31 in the 1930s, 14.75 in the 1940s, and 14.56 in the 1980s, while $\langle$V$\rangle$ ranges from 14.47 in 2000-2001 to 15.09 in 2005-2009.   However, no sustained or systematic trend are apparent from 1895 to 2009.

The pre-eruption brightness level has a number of important applications for the consideration of the evolution of novae and cataclysmic variables in general.  For these applications, it is likely best to use the average brightness level.  With this, the pre-eruption magnitudes are $B_{pre}$=14.55 from the Harvard plates and $V_{pre}$=14.62 from the {\it ASAS} magnitudes.  The $B$ magnitudes can be converted to $V$ magnitudes with the post-eruption $\langle$B-V$\rangle$=0.23.  The average magnitude depends on how the magnitudes are averaged.  The likely best method is to average the decadal averages, with this avoiding biases towards the level of years with large number of magnitudes.  With this, the mean pre-eruption magnitude is 14.5 with an uncertainty of roughly 0.2 mag.  The amplitude of the nova eruption ($V_{pre}$-$V_{peak}$) is 9.1$\pm$0.2 mag.  This is a rather low amplitude, pointing to a high mass accretion rate and pointing to an RN status.

\subsubsection{KT Eri Nova Eruptions Before 2009}

No prior event was discovered.  Our detailed search can be used to greatly limit the fraction of time from 1888 to 2009 over which any eruption could have happened without discovery.  We have archival data from every year that show KT Eri to not be in outburst.  During each observing season, we always have enough deep plates to prove that KT Eri could not have erupted.  But this still leaves the possibility that KT Eri erupted when it is inside the 'solar gap', that is, the seasonal interval around the time of its yearly conjunction with the Sun.  These solar gaps last for usually 4-6 months centred on May of each year.  

The solar gaps start with the last plate of each observing season (typically around March of each year) and ends with the first plate of the next observing season (typically around September of each year).  If KT Eri had a nova eruption that peaked one day after the start of the solar gap, then the last plate of the  prior observing season would not have detected any eruption.  However, it is possible that such an eruption would be detected in the tail of its eruption by the first plate taken at the end of the solar gap.  For example, if the next observing season started with a photographic plate with a limiting magnitude of $B$=14.5 with no KT Eri, than any eruption must have been $>$460 days earlier, and this proves that no undiscovered eruption could have popped off during the solar gap.  Important for this analysis is the strong result that all known RN eruptions from the same star have essentially identical light curves (Schaefer 2010).  If the first photograph of an observing season has a relatively poor limit, say only that $B$$>$12.0, then the peak of the undiscovered KT Eri eruption would have to be more than 163 days earlier, and this might not fill the solar gap if its duration is longer than 163 days.  In general, each year's solar gap might have effective coverage with any nova event during the gap being discovered, or there might be a short interval at the start of the solar gap during which KT Eri might have peaked yet producing no evidence of the eruption in the archival record.  These situations are illustrated in Fig. 3 for the decade of the 1950s.

\begin{figure}
	\includegraphics[width=1.1\columnwidth]{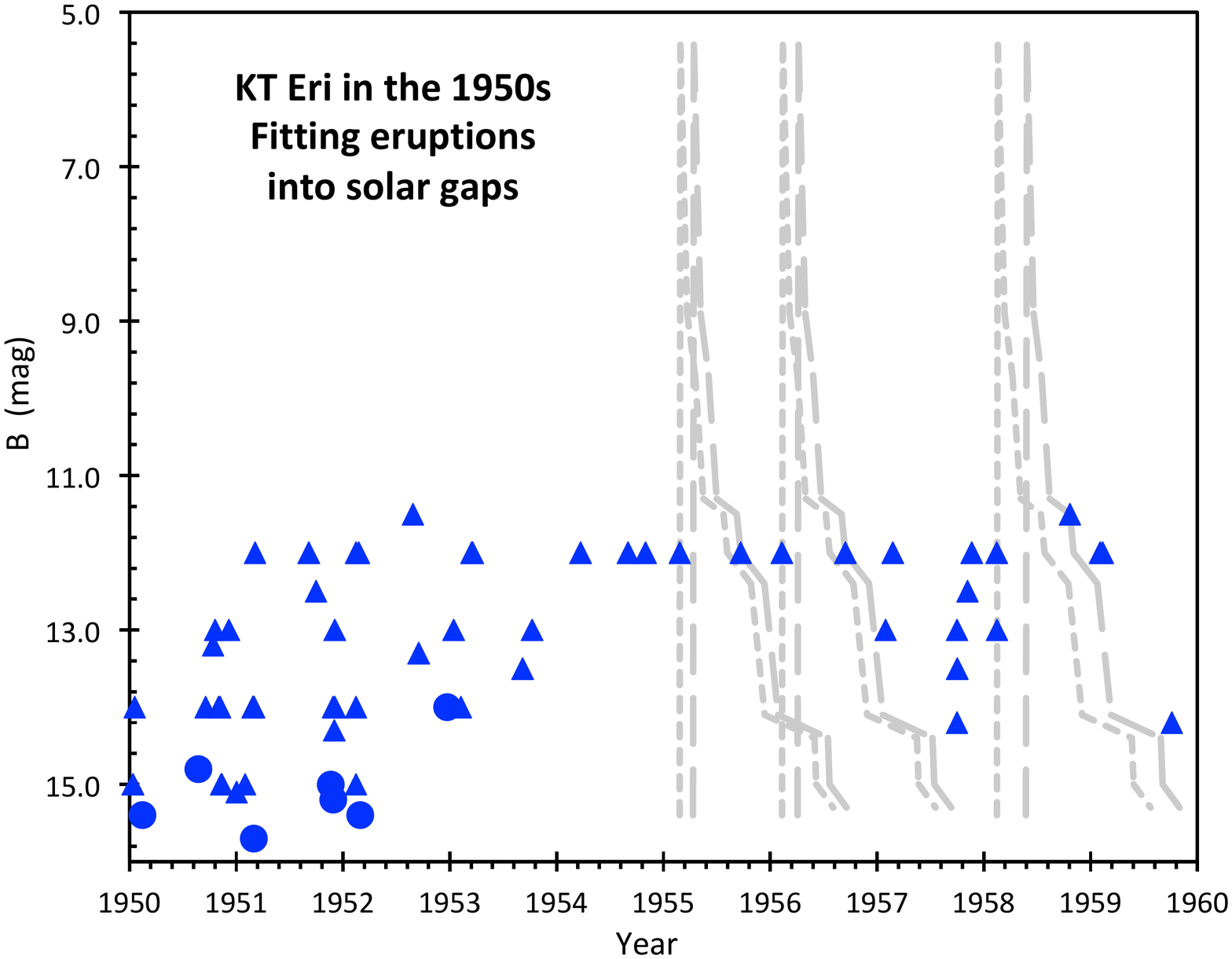}
    \caption{KT Eri in the 1950s.  Archival photographic plates from Harvard and Sonneberg were examined for any prior eruptions, but none was found.  Plotted here are the recorded Johnson-B magnitudes for positive detections of KT Eri in quiescence (blue circles) and for upper limits (blue triangles).  Solar gaps in our coverage (due to the Sun passing by Eridanus) extend typically for 4--6 months centred on May of each year.  For all the years from 1927--1954, KT Eri could not have erupted anytime within the solar gap without having been positively observed far above its quiescent level by the first image taken at the end of the gap.  But in some years, the first plate at the start of the next observing season is either relatively late or not particularly deep, and a KT Eri nova event could slip into the start of the solar gap and remain undiscovered.  In the 1950s, there is an interval of 46 days in 1955, 54 days in 1956, and 100 days in 1958 during which KT Eri could possibly have peaked and the eruption would remain undiscovered.  The plot shows the 2009 B-band light curve as a template for fitting into a gap, with the short-dashed gray curves showing the earliest possible peak in the three gaps, and with the long-dashed light curve showing the latest possible peaks for the three gaps.  The point of this plot is to show how the archival data can be used to sharply constrain possible dates for prior eruptions.}  
\end{figure}

We have tabulated the lengths of the solar gaps given all of our collected archival data for KT Eri.  For the majority of years, the gap has been closed.  That is, observations demonstrate that any eruption peak inside the solar gap would have been detected by the plate seeing KT Eri in the tail of the eruption as far brighter than the quiescent level.  Some of these remaining gaps are short, while the longest gaps are 297 days starting in 1894, 349 days starting in 1997, and 292 days starting in 2004.  From 1928 to 1954, there are no gaps at all, and this will become critical for our analysis and conclusions.  From 1888 to 2021, the total number of gap days is 2542, which is 5.2 per cent of the time.  The number of uncovered days (and the fraction of the total time) is given in Table 4 for a number of year intervals.

\begin{table}
	\centering
	\caption{Number of days during which KT Eri might have peaked in an undiscovered eruption}
	\begin{tabular}{llr} 
		\hline
		Year range & Days not covered & Fraction  \\
		\hline
1888--1908	&	733	&	10.0\%	\\
1909--1927	&	586	&	8.9\%	\\
1928--1954	&	0	&	0.0\%	\\
1955--1969	&	325	&	6.4\%	\\
1970--1999	&	497	&	4.7\%	\\
2000--2009	&	400	&	12.2\%	\\
2010--2021	&	0	&	0.0\%	\\
		\hline
	\end{tabular}
\end{table}

If KT Eri is an RN, then its recurrence time ($\tau_{rec}$) must be less than 100 years.{\footnote{This threshold of one century is the long-used traditional value.  But there is no physics reason to chose any specific dividing line between CNe and RNe, as the systems form a continuum of recurrence times.  One century is a nice round number that is appropriate for the observational situation of usually not having more than a hundred years of time for detecting nova events from any single system.}}  From Table 4, we could have missed an old eruption from 1909--1927, for 82$<$$\tau_{rec}$$<$100 years.  There is no possibility that any old eruption could have peak in 1928--1954, hence $\tau_{rec}$ cannot be from 55 to 82 years.  This archival constraint allow $\tau_{rec}$$<$55 years.

Within the RN hypothesis, any $\tau_{rec}$$<$40 years or so would require {\it three} or more missed eruptions.  For example, a 40 year recurrence time-scale would require undiscovered eruptions around the years 1889, 1929, and 1969.  In this case, the 1889 eruption could easily have been missed, but the joint probability will be less than 1 per cent that all three events will fall into coverage gaps.  (Indeed, for the case of a 40 year recurrence time-scale, the 1929 eruption would have to have come at least two years earlier to avoid the 1927--1954 interval with no eruption possible.)  The Galactic RNe are consistent with something like a 20 per cent jitter in eruption intervals (Schaefer 2010), with the sole exception of T Pyx with a large secular trend for slowing down (Schaefer et al. 2013).  For a 30 year recurrence time-scale, the undiscovered eruptions would have been around the years 1889, 1919, 1949, and 1979.  In this case, the 1889 eruption could easily have been missed, but the 1919 and 1979 eruptions should likely have been discovered, and the 1949 eruption would have certainly been caught.  With three-or-more lost eruptions, the probability will be greatly less than 1 per cent that all the eruptions just happened to peak inside one of our coverage gaps.  The possibility that $\tau_{rec}$ is less than $\approx$40 years can be rejected as requiring an unacceptably-low probability in the eruption times so as to avoid any of the prior eruptions being discovered.


\subsubsection{Post-Eruption Light Curve}

The post-eruption era starts when the light curve has stopped its decline and gone flat, with this certainly after the middle of 2011.  We have collected magnitudes from our own SMARTS program, from the AAVSO data base, from the APASS data base with time stamps, and from Munari \& Dallaporta (2014), see Tables 1 and 2.  The V-band magnitudes are plotted in Figure 4.

\begin{figure}
	\includegraphics[width=1.05\columnwidth]{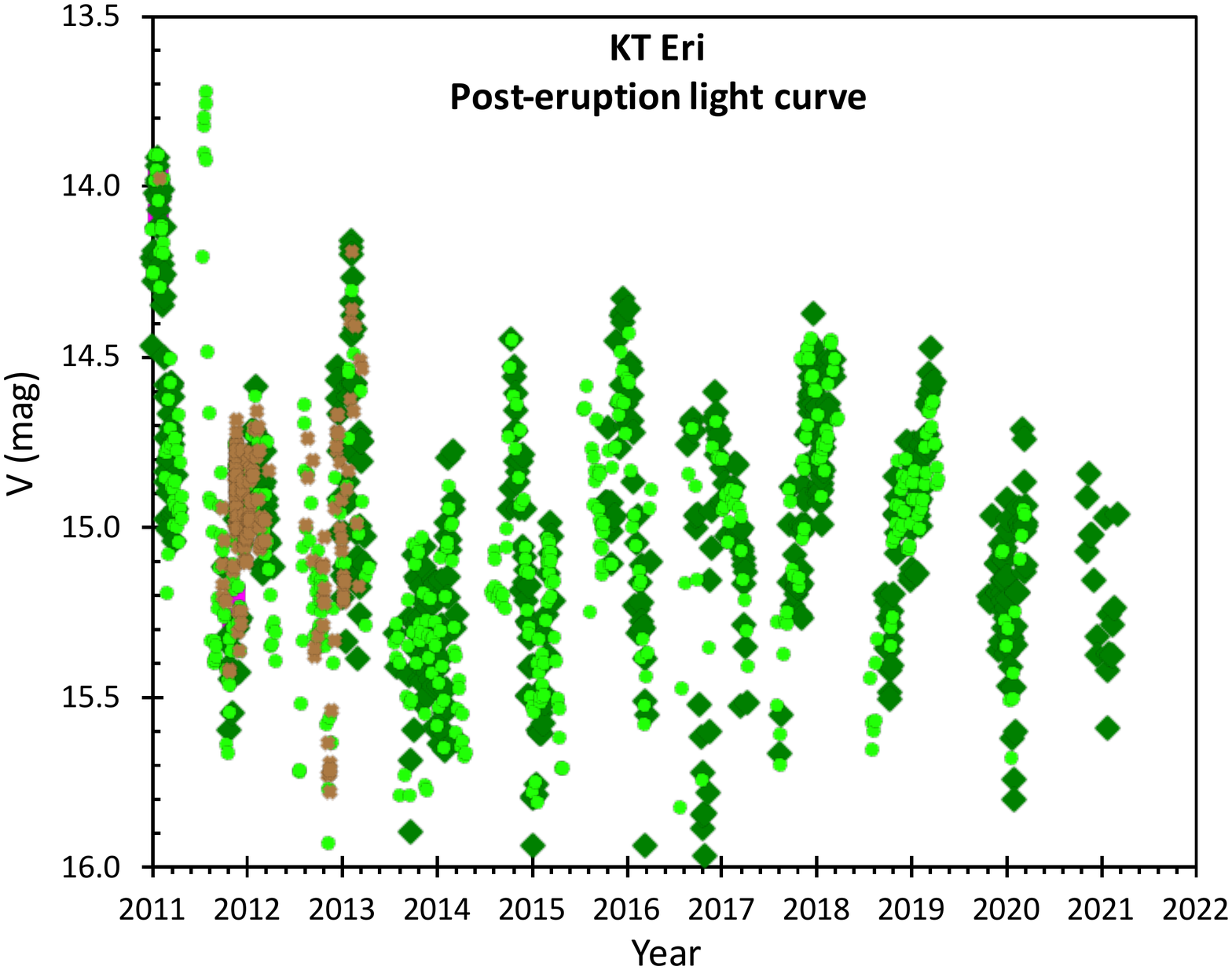}
    \caption{Post-eruption V-band light curve for KT Eri.  Our SMARTS magnitudes are displayed with light green circles, the AAVSO magnitudes are shown with dark green diamonds, and the V-band magnitudes from Munari \& Dallaporta (2014) are in brown crosses.  The eruption light curve is certainly over by mid-2011.  One of the points to take from this diagram is that KT Eri displays variability by over 1.7 mags, with frequent swings in brightness by over one magnitude on a time-scale of 1--3 months.  Another point from this diagram is that there exists no long photometric periodicity after the eruption is over, as confirmed with power spectra from discrete Fourier transforms.}  
\end{figure}

\begin{table}
	\centering
	\caption{SMARTS colors for KT Eri in post-eruption quiescence}
	\begin{tabular}{lrrr} 
		\hline
		Color  & Faintest & Average & Brightest  \\
		\hline
$\langle V\rangle$	&	15.84	&	15.06	&	13.88	\\
$\langle B-V\rangle$	&	0.31	&	0.23	&	0.20	\\
$\langle V-R\rangle$	&	0.33	&	0.25	&	0.14	\\
$\langle R-I\rangle$	&	0.34	&	0.27	&	0.18	\\
$\langle I-J\rangle$	&	0.13	&	0.05	&	0.07	\\
$\langle J-H\rangle$	&	0.43	&	0.28	&	0.16	\\
$\langle H-K\rangle$	&	-0.08	&	-0.05	&	-0.01	\\
		\hline
	\end{tabular}
\end{table}

The light curve is startling for its large amplitude of variations.  The total range of variations is over 1.7 mag, with many instances of where the brightness steadily increases or decreases by over 1.0 mag over durations from 1 to 3 months.  To the best of our knowledge, this extreme variability is unprecedented for any nova in quiescence.

For purposes of calculating absolute magnitude for later use, there are a range of post-eruption magnitudes that we can chose from.  The average $V$-magnitude after 2011.5 is 15.06 mag.  The range of post-eruption $V$-magnitudes is roughly 13.9 to 15.9.  The associated colors are presented in Table 5 for the 9 faintest nights, for the decadal average, and for the seven brightest nights.  The RMS scatter is close to 0.05 mag for all the colors.  When KT Eri brightens, its light gets slightly bluer, as appropriate for adding light from the accretion disc that is bluer than the light from the companion star.

The post-eruption light curve can be used to test prior period claims and to search for new possibilities.  For example, Munari \& Dallaporta (2014) report a 752 day periodicity of apparently eclipse-like minima.  This was based on the pre-eruption light curve of Jurdana-\v{S}epi\'{c} et al. (2012) plus just one post-eruption `eclipse' (in 2012.87).  Unfortunately, their pre-eruption light curve was not reproduced by our measures, many of the claimed `eclipse' times display apparent flares, and no `eclipses' are seen.  Further, the predicted eclipses of 2016.98 and 2019.04 are at times of maximum light, while the predicted eclipse for 2014.92 misses a wrong-shaped minimum by two months.  That is, the period of 752 days fails badly.

We have searched for any plausible periodicity from 0.1 days to 5 years in the post-eruption data.  The extreme variability appears as flares and dips with time-scales of 1--6 months.  Such dips and flares can provide alluring possibilities for periodicities to appear to the human pattern-recognition ability.  And indeed, a discrete Fourier transform does show high and isolated peaks.  The highest peak is for a period of 436 days.  But this peak is just created by the beating of the three most prominent dips (each with much photometry).  With the random alignment of these three dips, the folded light curve looks like a noisy sine wave.  But this best periodicity fails completely because most of the other times of minima are certainly not near any observed minima, while one is well centred on the highest flare.  What is going on is that a light curve with a modest number of random flares (or dips) will always have sharp and high peaks in the Fourier transform caused by random alignments of the flare times.  In the context of Gamma Ray Bursts, Schaefer \& Desai (1988) calculate the probabilities that shot noise will produce spurious periodicities, with this formalism applicable for the post-eruption flares of KT Eri.  The result is that even the best periodicities for the post-eruption data are insignificant, merely being the result of pattern recognition in random flare dates.

We have also looked for photometric periods from 0.04 days to 5 days (the full range of nova orbital periods, excluding those four RNe with red giant companions).  No such periodicity was found.  The noise confusion limit starts for a full amplitude of 0.12 mag.  In all, no significant period was found for KT Eri. 

\subsubsection{{\it TESS} Light Curve}

The {\it Transiting Exoplanet Survey Satellite} ({\it TESS}) covers 75--80 per cent of the sky down to roughly 19th mag, where all covered stars have generated light curves with 20 to 1800 second time resolution, all continuously for 26-day intervals.  This is perfect for finding orbital periods of novae (Schaefer 2021b).

{\it TESS} observed KT Eri for two separate intervals, called Sectors.  Sector 5 started on 2018 November 15, with a duration of 26.6 days.  For this Sector, KT Eri was not targeted, and hence the only data available are the Full Frame Images, taken continuously with exposures of 1800 seconds.  Sector 32 started on 2020 November 20 with a duration of 26.0 days.  For this Sector, KT Eri was targeted, which means that the small images around the target (called Target Pixel Files) were read out and saved every 20 seconds, all as a non-stop time series for 26.0 days, although with a one-day gap in the middle between orbits.  This is good for seeking both orbital and spin periods.

We went from the primary {\it TESS} data to the final light curves by using standard tools and processing pipelines (Jenkins et al. 2016).  Much of this was run with the {\sc Lightkurve} program (Lightkurve Collaboration, 2018).  All of the data and analysis programs are freely available at MAST{\footnote{https://mast.stsci.edu/portal/Mashup/Clients/Mast/Portal.html}}.  In essence, this is simply to perform aperture photometry on KT Eri for each image.  These images for photometry are small Target Pixel Files when extracted every 20 seconds onboard the satellite during Sector 32 or extracted from the Full Frame Images downloaded to the ground every 1800 seconds during Sector 5.  The {\it TESS} pixels are squares 21 arc-seconds on a side, where a centered pixel will ensquare 50 per cent of the flux.  We have used a 3$\times$3 array for our photometry aperture as this includes typically 99 per cent the starlight.  For the case of KT Eri, no nearby stars provide significant extra flux into the photometry aperture.  The background is determined from star-less regions also recorded in each Target Pixel File.  This background does change substantially throughout each spacecraft orbit of 13.7 days, due to scattered light.  The reported flux, in units of electrons per second, is simply the flux in the aperture minus the scaled flux for the background.  With the stability of the spacecraft environment, this leads to very good photometric precision.  Nevertheless, small effects still arise, typically at the 0.001 mag level, although sometimes the uncertainties can be substantially higher.  These effects are imperfect background subtraction and small spacecraft pointing drifts that change the fraction of the target light inside the photometry aperture.  Further, we have rejected a small fraction of the fluxes for cause, including single magnitudes flagged as being of lower quality (e.g., exposures during a spacecraft `momentum dump'), and brief intervals often at the start or end of orbits where the flux of all the stars changes systematically.  There are several standard tools for correcting for these small instrumental artifacts, and in the case of KT Eri, these are in good agreement.  The fully-corrected fluxes are labelled as `pdcsap\_flux' with units of electrons per second.  The spectral sensitivity of the {\it TESS} CCDs runs from 6000--10000~\AA.  The mid-exposure times are Barycentric Julian Dates (BJD).

For Sector 32, the full-resolution 20-second exposure light curve has 102757 fluxes from BJD 2459174.23289 to 2459200.11816.  With this full-resolution, the light curve follows a smooth curve with superposed flares, where the scatter arises from ordinary Poisson variations.  The light curve shows ordinary flickering on time-scales from minutes to days.  

The light curve for Sector 32 is shown in Figure 5.  For display purposes, the light curve is shown where the 20-second exposures are averaged together into bins of size 0.02 days.  The reason is that this beats down the ordinary Poisson variations, all while not hurting the visibility of any orbital modulation.  This same binned light curve is placed into Table 2, to avoid swamping the file with 102757 lines.

\begin{figure}
	\includegraphics[width=1.00\columnwidth]{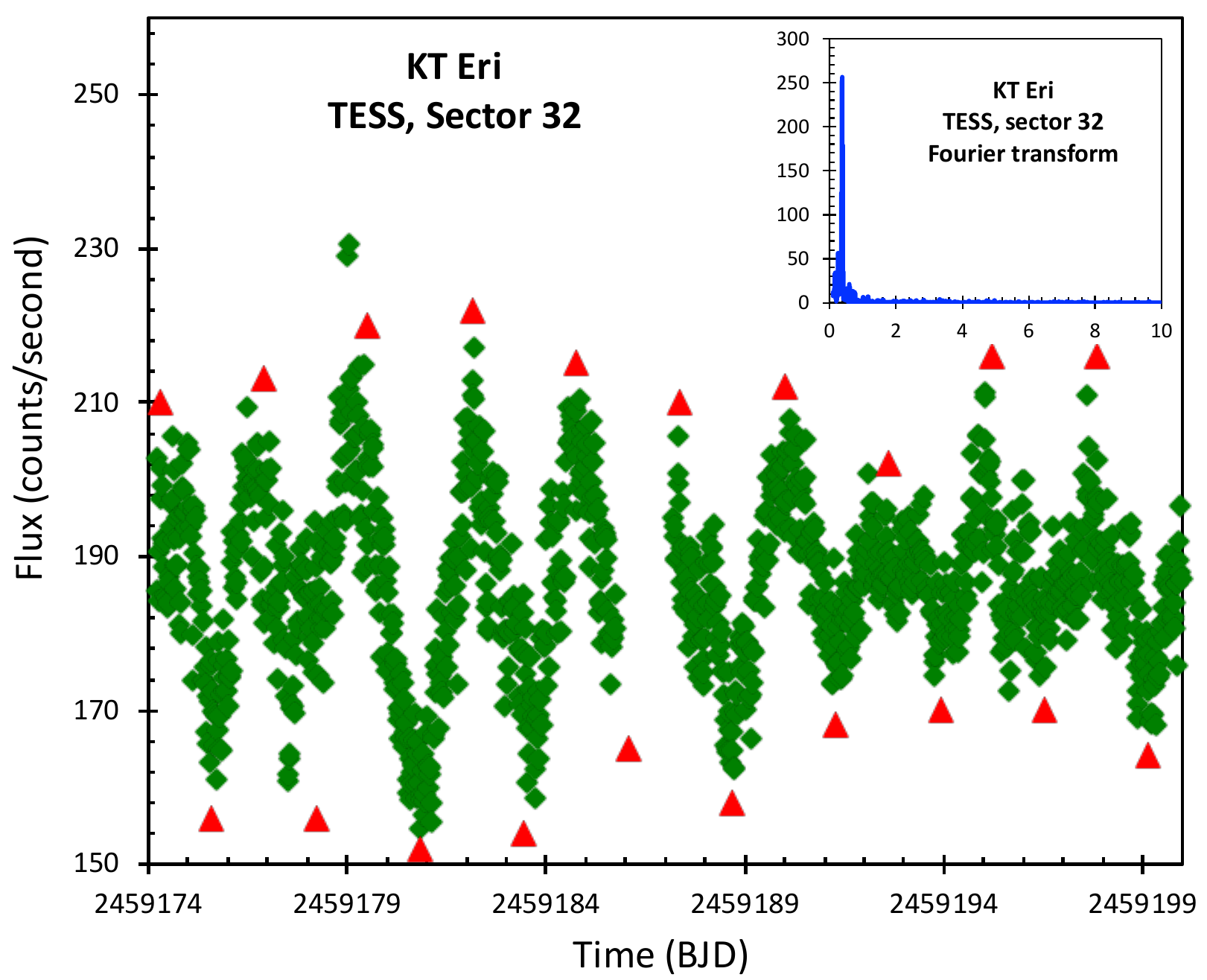}
    \caption{{\it TESS} light curve for Sector 32.  Importantly, this Sector shows a prominent 2.6 day periodicity.  The period is coherent, as can be seen by the maxima and minima all lining up with the red-triangular time markers that are spaced at 2.61595 days.  (This is the accurate period from the radial velocity curve.  However, this period is not accurate enough to produce reliable phases across the intervening 9 years.)  The bottom time markers are for the minima, while the top markers are offset by half the period and they follow the times of the light curve peaks.  The shape of the peaks looks to be triangular, but the folded light curve shows an apparent sine wave.  The amplitude of the orbital photometric modulations varies by roughly a factor of 4$\times$.  The periodicity is also prominent as an isolated very-high peak in the Fourier transform, as shown in the inset panel which plots power in units of the average power versus the frequency in units of cycles/day.  This provides precedent that KT Eri has a large and intrinsic change in amplitude on a time-scale of days to weeks.}
\end{figure}

With the full resolution light curve, we have sought coherent photometric periodicities with Fourier transforms, as shown in the inset in Fig. 5.  The period range of our full search was from 30 seconds up to 10 days.  Only one significant period was found, near 2.6 days.  This period is very highly significant.  The power spectrum is displayed in the inset to Fig. 5.  This period is prominent and easy to see in the data (Fig. 5).  KT Eri has its maxima and minima faithfully following the ephemeris for a 2.6 day periodicity.  There is no doubt about the significance and existence of this periodicity.  With a photometric period of 2.6 days and its coherence, this can only be the orbital period.

What is the uncertainty in the orbital period?  The formal calculations for four different methods (a chi-square fit to a sine wave, a Fourier transform, and two periodogram methods) all give error bars of $<$0.00035 days, despite these four methods returning $P$ values over a range of 0.021 days.  These error bars are greatly too small because the methods assume that the Gaussian noise from point-to-point is uncorrelated, whereas the {\it TESS} light curve has the noise around the average orbital modulation dominated by many large flares (i.e., flickering).  What is happening is that the flickering will shift the apparent times of minima and maxima both earlier and later, depending on the random timing, by $\sim$10 per cent of the period.    Schaefer (2021a) proved with {\it TESS} light curves for four cataclysmic variables that the ubiquitous jitter in the minimum-to-minimum times is caused by this random flickering causing random jitter of individual minimum times.  We know of no formalism or calculation that can give the period error bars in the presence of flickering.  But we can make an approximate estimate.  Considering the first and last minima, with each uncertain by $\sim$10 per cent of the period, the time difference will be uncertain by $\sim$14 per cent, or $\sim$0.37 days for KT Eri.  With Sector 32 having 10 cycles, the uncertainty in $P$ will be $\sim$0.04 days.  We do not know which of the four methods is best, all we can do is take some middle value, which is near 2.64 days.  In all, Sector 32 has a realistic period estimate of 2.64$\pm$0.04 days.

The Sector 32 light curve shows a prominent periodicity with a constant period, but the {\it amplitude} of the modulation apparently changes significantly.  The light curve swings up and down by roughly 40 counts/second round the third cycle, whereas this amplitude is roughly 10 counts/second in the eighth cycle.  It looks like some sort of a beating phenomenon where the amplitude varies at a beat frequency.  This variable amplitude of KT Eri has good precedents, as described below.
 
The light curve for {\it TESS} Sector 5 began on BJD 2458440.5154 and ended on BJD 2458463.9114.  This time series has exposure times of 1800 seconds, and there are 1050 fluxes.  These are plotted in Figure 6, and listed in Table 2.  

\begin{figure}
	\includegraphics[width=1.00\columnwidth]{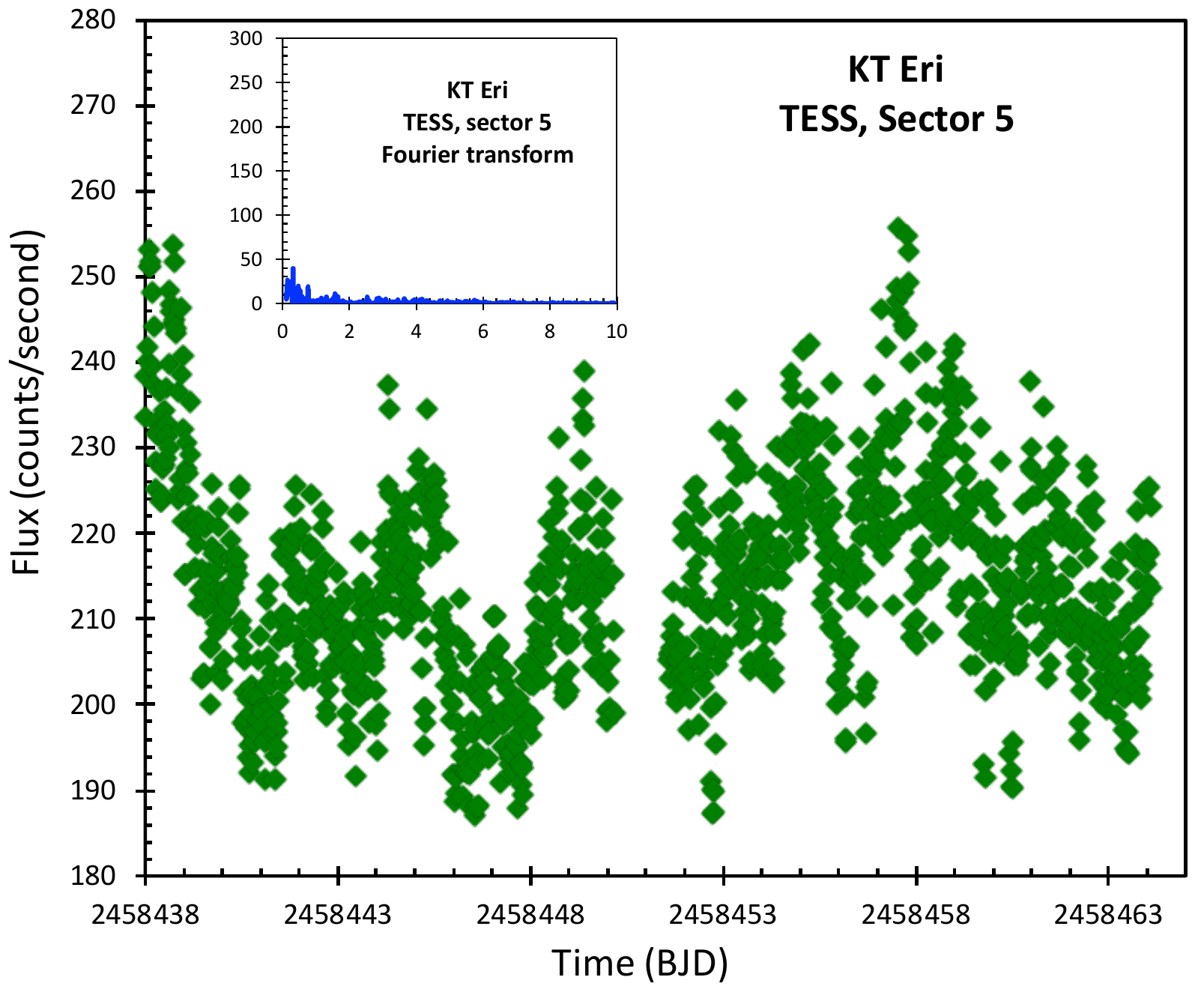}
    \caption{{\it TESS} light curve for Sector 5.  Importantly, this Sector shows no significant periodicity, not for 2.6 days nor any other period.  This can be seen in the inset panel showing the Fourier transform, with the same scale as the inset for TESS 32 in Fig. 5.}  
\end{figure}

For both sectors, the formal 1-sigma measurement error is close to 1.4 counts/second (derived from the propagation of the Poisson uncertainties), with this being substantially smaller than the point-to-point scatter in the light curve.  We know of no systematic problem or further measurement error that operates from point-to-point, hence the observed scatter in the light curve can only be intrinsic to the star.  That is, on time-scales of faster than half-an-hour, KT Eri is fast flickering with a RMS amplitude of 6 count/second (0.03 mag).

The Sector 5 light curves shows a number of small amplitude ups and downs with time-scales of a day or so.  It is natural to look at the relative spacing of the flares and dips and try to form a periodic progression.  But these maxima and minima do {\it not} follow anything like a 2.64 day period.  And these maxima and minima do not form any apparent periodicity that is sustained throughout Sector 5.  A Fourier transform (see inset in Fig. 6) does, of course, have a highest peak, and for Sector 5 it is at 3.23 days.  But this peak has a maximum power of only 39, which translates into a insignificant period.  (This can be compared to Sector 32, whose peak rose to a power of 256 for a similar time resolution.)  The insignificance is emphasized by the fact that neither of the two orbits (halves) of Sector 5 produce a Fourier peak anywhere near to 3.23 days.  To further emphasize the insignificance of this `best' period, such modulations are not apparent in the ups and downs of the light curve in Figure 6.  So, Sector 5 is not showing the 2.64 day period, nor any other period.

\subsubsection{Variable Amplitude}

Superposed on the fast flickering, both sectors show ups and downs on days-long time scales.  In Sector 5, the average flux is 214 counts/second, and the full amplitude of variation is roughly 30 counts/second (0.14 mag), all with no significant periodicity.  The formal amplitude limit at a period of 2.6 days is close to zero, while even for a range of nearby periods, the signal is $<$3 counts/second ($<$0.014 mag for full amplitude).  In Sector 32, the average flux is 187 counts/second, with variations whose amplitude varies from 10 to 40 counts/second (0.06 to 0.24 mag).   These variations are periodic with $P$=2.64 days, and the amplitude appears to vary systematically in a manner that looks like it could arise from some beating between two close periods. Why are there these stark differences in variability amplitude between Sectors 5 and 32?

The changing amplitude cannot be a simple beating phenomenon.  The reason is that any ordinary interplay between two periods must have the duration of near zero amplitude always at a constant.  But Sector 32 shows a low amplitude for only a few days, whereas Sector 5 shows a low amplitude for at least 23 days.

Could the changing amplitude arise from some extra light (say, from the disc) being added in to swamp the periodic signal?  Sector 5 is brighter on average than Sector 32 by 25 counts/second, however the upper limit on the periodic signal ($<$1.3 counts/second) is greatly below the range of periodic signals from Sector 32 (10 to 40 counts/second). Adding extra light does not explain the low amplitude in Sector 5.

To explain the low periodic amplitude in Sector 5, similar systems can provide precedents.  We know of only five novae that display highly variable amplitude in their orbital modulations, and these are all with similar binary systems.  These five nova systems are in the very narrow class of novae with subgiant companions with orbital periods between 1.2 and 3.5 days, many of which are RNe.  {\bf (1)} The RN V394 CrA has periodic modulations on the orbital period (a well-measured 1.5 days) that occasionally go away{{\footnote {We have extracted the {\it TESS} light curve for Sector 13, and prominent eclipses appear with a period of 1.5 days.  The amplitude goes from 0.045 mag on the first day, steadily decreasing in amplitude for the next six orbits until there is no visible modulation at all for the next 12 days, only to have the minima reappear with increasing depth for the last three orbits.}}.  This was known from extensive photometry over many years from Cerro Tololo, with the variable amplitude being perplexing.  Even in the old days, it was recognized that the amplitude was largest when V394 CrA was faintest, and the dips went away only when the star was near its brightest (Schaefer 2010).  V394 CrA provides a perfect precedent for an RN with an orbit a bit longer than one-day having starkly variable amplitudes, with 12-day-long intervals of zero amplitude, all tied to the accretion rate, with the zero-amplitude intervals when the accretion rate is highest.  {\bf (2)} Our second precedent is the recurrent nova V2487 Oph (Schaefer, Pagnotta, \& Zoppelt 2022).  We observed this with the {\it K2} mission of the {\it Kepler} satellite and discovered an orbital period of 1.24 days.  Over the 69 day observing interval, the amplitude started out at a middle level, increased steadily to day 15 (with amplitude near 0.08 mag), then steadily decreased in amplitude to day 40, with a zero amplitude from day 40 until day 53, whereupon the amplitude rose to near the maximum by the end of the run on day 69.  V2487 Oph is a perfect precedent for a recurrent nova with a period just longer than one-day for which the orbital modulations varied in amplitude from 0.08 mag down to zero, with intervals of zero amplitude lasting for two weeks.  {\bf (3)} Our third precedent is the recurrent nova U Sco with a very well measured orbital period of 1.23 days and deep eclipses.  During the late tail of the 2010 eruption, when the disc was reforming, deep intermittent `eclipses' were seen at all orbital phases (Schaefer et al. 2010).  This was interpreted as the fast varying splash from the accretion stream making a transient high outer edge of the disc which caused the eclipse of the inner bright region near the white dwarf.  U Sco is a precedent and prior realization that an RN, with orbital period just longer than one-day, has a variable height of the outer edge of the disc, making for rapidly changing amplitude of the orbital modulation.  {\bf (4)} Our fourth precedent, and the only non-RN system, is V407 Lup, a nova from 2016.  The {\it TESS} light curves from Sector 12 and 38 have a period of 3.57 days (Schaefer 2021b).  Inside both sectors, the amplitude appears to vary substantially, although with relatively few orbits in each Sector, this is difficult to be exacting.  Between the two Sectors, the amplitude changes from 2.6 to 0.8 counts/second.  V407 Lup is not an RN because its accretion rate is too low.  {\bf (5)} Our fifth precedent for the Sector 5 orbital amplitude being zero is KT Eri itself, in Sector 32.  That is, KT Eri has already displayed amplitude changes from 0.24 mag down to 0.06 mag.  KT Eri itself provides the precedent that such systems have highly variable photometric oscillations on the orbital period, and it is not surprising that Sector 5 can show an interval with zero amplitude.

This is prior and strong precedent that a very narrowly defined class of nova systems has large variations in the amplitudes of their orbital modulations.  These precedents are all for nova systems with orbital periods from 1.2 to 3.5 days, and a reasonable idea would be that the conditions necessarily have a narrow range of disc sizes over which the effects are observable.  We know of no cases of large changes in amplitude outside of this very narrow range of periods.  However, for nova systems inside the 1.2$<$$P$$<$3.5 day range, V2109 Oph, and V392 Per (Schaefer 2021b), as well as GK Per do not display variable amplitudes.

\subsection{Spectral Energy Distribution}

Munari \& Dallaporta (2014) report a spectral energy distribution (SED) for KT Eri in quiescence.  Their seven input magnitudes were their own post-eruption BVRI measures plus the pre-eruption JHK measures from the {\it 2MASS} survey.  They also point out that the {\it WISE} infrared magnitudes showed KT Eri only during the tail of the eruption, and these points could not be added to the SED for quiescence.  Their SED plot shows a power law with modest curvature.  They fit the SED as being that of an A7 {\rm III} star at 5 kpc distance.  This fit has three major problems:  First, the optical and infrared parts of the SED were taken a decade apart, and KT Eri varies greatly by up to two magnitudes during quiescence, hence their overall shape of the SED is inevitably skewed with a likely-large offset from the optical to infrared in the reported SED.  Second, the spectral model includes no disc contribution to the SED, yet the disc component is very bright, dominating over the companion star.  Third, their SED covers the frequency region where the disc light is switching from the $F_{\nu} \propto \nu^{1/3}$ slope to the Rayleigh-Jeans region with a slope for $F_{\nu} \propto \nu^2$, while the disc spectrum shows curvature that looks much like that displayed in Figure 3 of Munari \& Dallaporta (2014).  That is, their fit is not measuring the temperature of a blackbody-like companion star, but rather is telling us about the accretion disc.

We have constructed our own SEDs, with all three of these problems being solved.  Critically, we have complete sets of $BVRIJHK$ photometry from 272 nights with KT Eri in quiescence (from July 2011 to April 2019).  All magnitudes from each night were taken within a ten minute interval, indeed with many pairs of magnitudes taken simultaneously.  With this, we have constructed SEDs for many individual instances, for seasonal averages, for KT Eri at its brightest, and for KT Eri at its faintest.  With the averaging of many nights data together, the photometric error bars become small, allowing for good spectral model fits.  Further, we have made model fits as a realistic $\alpha$-disc plus a companion star.

Our SEDs are constructed from the SMARTS data set, with the {\it ANDICAM} instrument on the Cerro Tololo 1.3-m telescope.  ({\it ANDICAM} uses a dichroic mirror such that the optical and infrared images are taken simultaneously.)  The RMS scatter of magnitudes for seasonal averages in all of the bands is typically one-third of a magnitude, hence the uncertainties on the average magnitudes are usually around 0.05 mag.  The conversion from magnitudes to fluxes (in units of Jansky) uses the calibrations adopted in Schaefer (2010).  The extinction correction was made with $E(B-V)$=0.08 mag, and hence $A_V$=0.25 mag, as taken from Ragan et al. (2009).  Many of our SEDs are plotted in Fig. 7.

\begin{figure}
	\includegraphics[width=1.1\columnwidth]{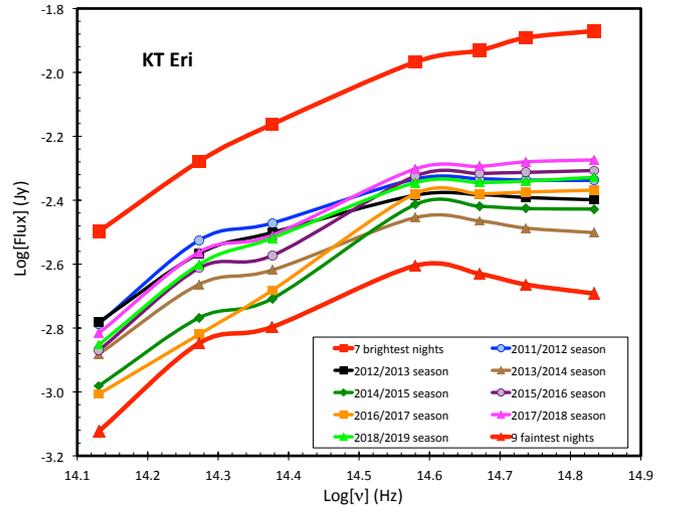}
    \caption{KT Eri spectral energy distribution in post-eruption quiescence.  These SEDs show the flux levels ($F_{\nu}$ in Janskys) as a function of the central photon frequencies ($\nu$ in Hertz) for the K, H, J, I, R, V, and B bands (from left to right).  These SEDs are the summation of the disc light plus the blackbody light from the companion star.  The top red curve shows the average SED for the 7 brightest nights (all with $V$$<$14.21), and the bottom red curve shows the average SED for the 9 faintest nights (all with $V$$>$15.77)  The middle curves show the seasonal average SEDs from 2011/2012 to 2018/2019.  KT Eri has an extreme amplitude of variation in quiescence.  The curvature and the peak near the R-band are from the blackbody of the companion star.  These SEDs are fit to a model with a standard $\alpha$-disc model plus a blackbody for the companion star.  In the faint-state, the effective temperature is 6200$\pm$500 K.}  
\end{figure}

Our SEDs show the huge range of variation for KT Eri.  At the top is the SED for the 7 brightest nights (with $V$$<$14.21), for which the accretion disc dominates over the companion star.  At the bottom is the SED for the 9 faintest nights (with $V$$>$15.77), for which the blackbody spectrum of the companion star is seen on top of the power law of the accretion disc.  Between these extremes are 8 SEDs for the seasonal averages from 2011/2012, 2012/2013, ... , 2018/2019.  The shape of the SEDs has a steady progression from the companion-star-dominated faint state to the accretion-disc-dominated bright state.

Our SED  model consists of a disc component plus a companion star component.  The disc component is taken as the standard $\alpha$-disc model (Frank, King, \& Raine 2002).  For this, we have performed the full integration over blackbody annuli from the white dwarf surface to the outer edge of the disc near the Roche surface.  The primary unknown fitting parameter is the accretion rate, $\dot{M}$, but its use here is merely to get a realistic curvature in the model SED.  This calculated disc flux is then multiplied by an arbitrary normalization factor that absorbs many uncertainties (including the inclination and distance).  The companion star is represented by a simple blackbody of temperature $T$, with the normalization allowed to freely vary.  This model has four fitting parameters, the two normalization constants for the power law and the blackbody, plus the blackbody $T$ and the disc $\dot{M}$.  Our fitting procedure is to adopt a value for $T$ and $\dot{M}$ over a large grid of values, vary the two normalization factors until the chi-square is minimized, then the best grid position can be recognized.  Our use of the two arbitrary normalization factors means that the fit is to the {\it shape} of the SED by adding two distinct shapes (for the disc and for the companion), with this allowing the recognition of the relative sizes of the two components.  Further, this allows us to pick out the companion star's component and recognize the Wien peak and measure the surface temperature.

The model fit for KT Eri at its faintest will provide the best measure of $T$.  In this extreme state, the observed brightness has an average of $V$=15.84.  This best-fitting value is 6200 K.  The error bar is uncertain due to not knowing all the intrinsic variations, as well as possible variants on the disc model (like truncating the disc far inside the Roche surface), but it appears to be roughly $\pm$500$\degr$.  From B to I to K bands, the model places the companion star as 1.4$\times$, 3.1$\times$, and 2.5$\times$ the brightness of the disc.  With KT Eri in its extreme faintest state, the disc light only provides roughly one-third of the system light.  In the V-band, the correction from the observed magnitude to a disc-only magnitude is 1.25 mag.

An important point from the SED gives the fundamental nature of the companion star.  With $T$=6200$\pm$500 K, the companion is too hot to be a main sequence star in any cataclysmic variable.  For any such companion, the star must have already evolved off the main sequence, and the consequent orbital period must be larger than 0.5 days or so.

Further, the size of the companion can also be measured from the SED with a calculation of its blackbody radius.  For the {\it Gaia} distance and our SMARTS SED with KT Eri at its faintest, subtracting out the disc light modeled as an $\alpha$-disc, the radius is roughly 3 R$_{\odot}$.  This is the size of a subgiant star of the measured temperature, therefore, the companion star must have evolved far off the main sequence.

The SED for KT Eri in its average state is taken as that of the 2018/2019 observing season (see Fig. 7).  This middle SED has the observed V-magnitude of 15.00.  This SED was fit with the same model, with the $T$ and the blackbody normalization factor held constant with the fit for KT Eri at its faintest.  We got a surprisingly close fit by only varying the $\dot{M}$, where the accretion apparently increased by a factor of 14$\times$ from the faintest state to the middle state.  By our fits adding up differently shaped components in the SED, we can distinguish what fraction of the light comes from the disc and companion as a function of the band.  From the B to I to K bands, the ratio of the fluxes from the companion star to the accretion disc is 0.23$\times$, 0.57$\times$, and 0.80$\times$ respectively.  In its average middle state, KT Eri has the accretion disc dominating roughly by a factor of 2:1.  For critical use later, in the V-band, the correction from the observed magnitude to the disc-only magnitude is 0.34 mag. 

KT Eri in its extreme high-state is taken from the 7 brightest nights out of all our 272 post-eruption simultaneous SEDs.  For these nights, the average is $V$=13.88, without extinction correction.  This SED (of course with extinction correction) is displayed as the top red curve in Fig. 7.  Our model fit uses the parameters for the companion star as unchanged from the values derived for the extreme-faint state.  The disc parameters were freely varied until the sum of the two components matched the observed shape of the SED.  This allows us to distinguish the relative fluxes from the disc and the companion.  To match the {\it shape} of the SED, the disc component must dominate greatly over the companion star's blackbody.  Our specific best-fitting ratio of the fluxes from the companion star to the accretion disc is 0.09$\times$, 0.21$\times$, and 0.28$\times$ for the B, I, and K bands respectively.  That is, roughly the disc dominates over the companion by roughly 4:1.  In the V-band, the fraction of light in the disc makes for a difference in the observed magnitude and the disc-only magnitude of 0.14 mag.  This correction will become important later in this paper for use in determining the accretion rate.

\subsection{Radial Velocity Curve}

Munari, Mason, \& Valisa (2014) present 67 radial velocities (from 19 nights) for the He {\rm II} 4686 \AA ~line. The radial velocities vary from -259 to -38 km s$^{-1}$.  However, the velocities change greatly from exposure to exposure on short time scales. For example, on one night, the radial velocity changed by 65 km/s in 16 minutes.  This makes it likely that non-orbital velocity variations (from measurement error or intrinsic to the star) are at this level.  With this scatter being similar to the $K$ velocity and with only 19 times of measurements, no significant periodicity was found in a Fourier transform.  The power spectrum shows many low peaks, with many near the top power, and no period is significant.

We have also constructed a radial velocity curve for the He {\rm II} 4686 \AA ~line with our SMARTS spectra  (Walter et al. 2012). We have 109 measures from 108 nights from early 2010 to late 2011, all with the RC Spectrograph on the Cerro Tololo 1.5-m telescope, as reported in Table 6.  A Fourier transform shows a highly significant peak at a period of 2.61575 days.  (There is a peak with significantly lower power for the period 1.611084 days, with this being an artifact from the window function, where the two frequencies add up to exactly one-per-day.  Our highest peak corresponds exactly with the second highest peak in the fourth highest cluster of peaks in the Fourier transform of the Munari et al. data.)  A formal chi-square fit with a sinewave gives 2.61595$\pm$0.00060 days.

The folded radial velocity curve is consistent with a circular orbit, with the middle value ($\gamma$) equal to $-$142$\pm$5 km s$^{-1}$.  The epoch of crossing $\gamma$, going more positive, is JD 2455491.323$\pm$0.053, with this corresponding to the conjunction with the white dwarf in front of the companion star.  The K amplitude is 58.4$\pm$7.0 km s$^{-1}$.

\begin{table}
	\centering
	\caption{KT Eri radial velocity data for He {\rm II} 4686~\AA ~line}
	\begin{tabular}{lllll} 
		\hline
		HJD & Velocity (km/s) &   &  HJD  &   Velocity (km/s)   \\
		\hline
2455206.500	&	-116	&		&	2455511.650	&	-203	\\
2455210.565	&	-214	&		&	2455513.679	&	-163	\\
2455211.586	&	-156	&		&	2455515.695	&	-113	\\
2455212.531	&	-86	&		&	2455516.623	&	-230	\\
2455215.582	&	-69	&		&	2455522.587	&	-193	\\
2455217.551	&	-79	&		&	2455524.741	&	-195	\\
2455222.551	&	-69	&		&	2455527.574	&	-250	\\
2455223.553	&	-116	&		&	2455532.551	&	-193	\\
2455224.523	&	-71	&		&	2455537.546	&	-246	\\
2455228.521	&	-124	&		&	2455543.535	&	-138	\\
2455229.500	&	-190	&		&	2455544.542	&	-168	\\
2455232.526	&	-100	&		&	2455549.525	&	-125	\\
2455234.527	&	-161	&		&	2455550.525	&	-179	\\
2455245.500	&	-125	&		&	2455553.550	&	-183	\\
2455246.527	&	-33	&		&	2455560.558	&	-219	\\
2455249.518	&	-103	&		&	2455562.546	&	-97	\\
2455251.527	&	-37	&		&	2455591.531	&	-90	\\
2455255.509	&	-177	&		&	2455593.539	&	-95	\\
2455261.552	&	-44	&		&	2455597.550	&	-180	\\
2455264.521	&	26	&		&	2455602.527	&	-141	\\
2455265.514	&	-197	&		&	2455605.518	&	-183	\\
2455269.501	&	73	&		&	2455607.553	&	-104	\\
2455273.500	&	-36	&		&	2455609.597	&	-92	\\
2455276.494	&	-98	&		&	2455615.564	&	-115	\\
2455279.489	&	-131	&		&	2455616.518	&	-154	\\
2455279.500	&	-23	&		&	2455619.524	&	-96	\\
2455282.489	&	-16	&		&	2455623.526	&	-139	\\
2455285.484	&	-16	&		&	2455625.504	&	-79	\\
2455286.519	&	-191	&		&	2455626.539	&	-154	\\
2455287.481	&	-72	&		&	2455628.501	&	-95	\\
2455290.480	&	-78	&		&	2455633.496	&	-117	\\
2455299.491	&	-139	&		&	2455637.541	&	-180	\\
2455304.500	&	-171	&		&	2455641.489	&	-89	\\
2455312.468	&	-195	&		&	2455647.493	&	-143	\\
2455392.942	&	-129	&		&	2455651.489	&	-57	\\
2455396.930	&	-199	&		&	2455655.386	&	-200	\\
2455405.500	&	-155	&		&	2455764.932	&	-232	\\
2455411.917	&	-224	&		&	2455768.904	&	-148	\\
2455416.898	&	-218	&		&	2455772.902	&	-238	\\
2455425.867	&	-184	&		&	2455783.895	&	-209	\\
2455428.893	&	-141	&		&	2455785.861	&	-185	\\
2455431.836	&	-144	&		&	2455801.815	&	-246	\\
2455451.795	&	-217	&		&	2455818.816	&	-173	\\
2455456.764	&	-208	&		&	2455820.794	&	-233	\\
2455460.837	&	-152	&		&	2455824.804	&	-182	\\
2455471.721	&	-176	&		&	2455830.764	&	-257	\\
2455476.705	&	-177	&		&	2455834.712	&	-134	\\
2455481.748	&	-179	&		&	2455838.803	&	-299	\\
2455484.832	&	-179	&		&	2455840.784	&	-166	\\
2455488.676	&	-157	&		&	2455844.724	&	-97	\\
2455489.801	&	-141	&		&	2455854.717	&	-225	\\
2455499.743	&	-99	&		&	2455858.659	&	-104	\\
2455501.822	&	-109	&		&	2455864.707	&	-208	\\
2455507.776	&	-121	&		&	2455871.614	&	-54	\\
2455509.750	&	-145	&		&		&		\\
		\hline
	\end{tabular}

\end{table}

\section{Analysis}

\subsection{Orbital Period}

KT Eri has many published photometric periodicities.  These include 35.09 seconds, 0.0938 days, 56.7 days, 210 days, 376 days, 737 days, 750 days, and 752 days, as well as quasi-periodic oscillations at 3530, 3809, 6560, and 10440 seconds.  These claims were all dubious to begin with, and critically, all have been subsequently disproven with further data extending the light curves.  The long periods were derived from much the same data set, being variations and aliases of each other. and are suspiciously close to multiples of the primary sampling period of one year.  Critically, these periods are not reproduced when the same data source is independently examined with Fourier transforms, and the spectrum in quiescence does not show a red giant companion as required for such long periods.  The claimed quasi-periodic oscillations have the observations only covering 1.2--2.1 cycles, with just two maxima/minima recorded, and a periodicity was claimed.  We have used the {\it TESS} light curve to search for significant broad-band power over a wide range of frequencies, and no significant periodicity was found.  The reported X-ray periodicity of 35.09 seconds was never significant, even by the reckoning of the claimants, as {\it post facto} data selection of a fraction of the light curve was required to get the confidence level of the periodicity up to 90 per cent, i.e. a 1.6-sigma detection even without accounting for the selection.  We have searched the {\it TESS} light curve for periodicities or quasi-periodicities from 30--40 seconds, with none being found down to the noise limit of 0.0029 mag.  The 56.7 day claim was based on only 3.5 cycles of data, and the 210 day claim was reported despite the authors stating "this detection does not appear to have high significance".  KT Eri has a poor history for claimed periodicities.

For KT Eri, both {\it TESS} and radial velocity data have produced highly confident measures of the orbital period.  These independently agree that the orbital period is near 2.6 days.  The radial velocity curve ensures that the orbital period is not double the optical modulation period due to ellipsoidal effects.  The two measures of $P$ are consistent to within their uncertainties.  We adopt the more precise period from the radial velocity curve, $P$=2.61595$\pm$0.00060 days.

This new $P$ measure has a variety of deep implications:  {\bf (1)} One implication is that the companion star must be an evolved star as required to fill up its large Roche lobe.  For a white dwarf mass of 1.25 M$_{\odot}$ and a companion mass of 1.0 M$_{\odot}$, the Roche lobe radius is 3.7 R$_{\odot}$, while the two star centres in the binary are separated by 10.4 R$_{\odot}$.  This radius is too small for a red giant and too large for any of the usual low-mass main sequence companion stars.  A 6200 K star of that size would be a subgiant `recently' evolved off the main sequence.  {\bf (2)} A second implication of our new long-period is that we have no understanding of why KT Eri has such a high accretion rate.  Models of cataclysmic variable evolution (e.g., magnetic-braking) are not applicable to binaries with $P$$>$7 hours, and the $P$=2.6 days of KT Eri means that the dominant source of angular momentum loss in the binary is unknown.  Is the accretion driven by the angular momentum loss in the binary, or by the evolutionary expansion of the companion?  And why does KT Eri have such a high accretion rate, while other similar systems (e.g., GK Per) have greatly lower rates?  {\bf (3)} A third implication is to realize we have no detailed idea of the evolution of KT Eri.   Most nova systems have periods between 0.05 days and 0.5 days (with nearly-main-sequence companion stars), while no published model of cataclysmic variable evolution applies to systems with evolved companion stars.  Presumably, the system started out as an ordinary detached binary, the more massive star evolved to become a white dwarf, later the companion started evolving off the main sequence, the Roche lobe overflow started, hence reaching the situation seen today.  With the observed high $\dot{M}$, the current state of KT Eri cannot last for longer than a few million years.  The ultimate fate of KT Eri depends critically on whether $M_{WD}$ is increasing towards the Chandrasekhar limit.  {\bf (4)} A fourth implication is that a 2.6 day period makes KT Eri look a lot like an RN.  Amongst Galactic RNe, 80 per cent have unusually long orbital periods (Schaefer 2010), whilst only 13 per cent of CNe have periods longer than 0.6 days (Pagnotta \& Schaefer 2014).  (And we expect that many of the long-period CNe are simply RNe with the second eruptions in the prior century having been missed.)  On the face of it, this makes for a reasonably high expectation that KT Eri is a recurrent nova.

\subsection{Orbital Inclination}

The inclination of the binary's orbital plane  is needed for a variety of calculations.  Ribeiro et al. (2013) presented a model reproducing the fine structure in the emission line profiles during the eruption.  That is, the ejecta has some reasonable presumed symmetry tied to the orbital plane, with the orientation of the axis of symmetry likely pointing along the pole of the binary orbit.  Their morpho-kinematical study of the H$\alpha$ line profiles points to a dumbbell structure with a 4:1 axis ratio and an inclination angle of $58^{+6}_{-7}$ degrees.  This is a reasonable estimate, although it could be easy to dismiss this value as being produced by an arbitrary and simplistic model for the ejecta.

Further, three constraints can be placed on the inclination:  {\bf (1)} The first constraint is that no eclipses are seen in the {\it TESS} light curve.  For the expected stellar masses with an orbital period of 2.6 days, this translates into a limit that the inclination must be less than 69$\degr$.  For all plausible stellar masses, this limit varies by only 2$\degr$.  {\bf (2)} The second constraint is that the {\it TESS} light curve shows the orbital modulations to have a full amplitude of 0.115 mag.  Such an amplitude rules out the binary orbit appearing nearly face-on.  A calculation of the amplitude expected as a function of inclination would require a very detailed geometry model plus a full knowledge of the irradiation of the companion star.  We are not able to do this calculation, yet it is reasonable to realize that the constraint will be that the inclination must be larger than some threshold like 30$\degr$.  {\bf (3)} The third constraint is our radial velocity curve, with $K$=58.4$\pm$7.0 km s$^{-1}$ and P=2.615 days.  For the most likely values with $M_{WD}$= 1.25$\pm$0.03 M$_{\odot}$ (see Section 3.7) and $M_{comp}$= 1.0$\pm$0.2 M$_{\odot}$ (see Section 3.8), the inclination is then 41$\pm$9 degrees.

For purposes of adopting an inclination for use with the analysis of the radial velocity curve to constraint the stellar masses in Sections 3.7 and 3.8, we cannot use the third constraint.  For this purpose, the first two constraints are not useful, and best value to use is just the 58$^{+6}_{-7}$ degree inclination from Ribeiro et al. (2013).

For purposes of knowing the binary inclination to calculate the accretion rate, we are allowed to adopt our independent white dwarf mass of 1.25$\pm$0.03 M$_{\odot}$, and a plausible range of companion mass (0.8--1.2 M$_{\odot}$).  With two useful inclination measures, $58^{+6}_{-7}$ degrees and 41$\pm$9 degrees, we adopt a weighted average to get a final inclination of 52$\pm$5 degrees.

\subsection{Extinction}

The only reported extinction measure for KT Eri is $E(B-V)$=0.08 mag (Ragan et al. (2009).  Schlafly \& Finkbeiner (2011) report that the extinction all the way through our Milky Way galaxy along the line of sight to KT Eri has $E(B-V)$=0.0821$\pm$0.0027.  KT Eri is certainly at a distance far past essentially all our galaxy's dust, and hence this upper limit is really a direct measure.  With this, the extinction in the V-band is $A_V$=0.254$\pm$0.008 mag.

\subsection{Distance}

Before {\it Gaia}, the distance to KT Eri was poorly constrained.  The only basis published was the so-called Maximum-Magnitude Rate-of-Decline (MMRD) relation (e.g., Munari \& Dallaporta 2014), but these are well known to be wrong by up to nearly 10$\times$ in distance for similar systems (Schaefer 2018).  Fortunately,  the {\it Gaia} spacecraft{\footnote{https://archives.esac.esa.int/gaia}} has recently measured an accurate parallax.  The {\it Gaia} EDR3 data release ({\it Gaia} Collaboration 2021; Lindegren et al. 2021) quotes the parallax as $\pi$ = 0.196$\pm$0.024 milli-arcseconds.  With the usual simple calculation, the distance is $(1000 pc)/\pi$ with the parallax in milli-arseconds, for a distance of 5100$\pm$620 pc.  However, to best translate the parallax into a distance, we must use the `exponentially decreasing space density' (EDSD) priors in a Bayesian calculation prescribed by the {\it Gaia} Team (Luri et al. 2018).  This uses an exponential length, $L$, for which the vertical scale height of 150 pc (corrected for the galactic latitude) is appropriate for the disk population of cataclysmic variables (Patterson 1984), which gives $L$=280 pc (Schaefer 2018).  With this expected $L$ value pulling to closer distances, the EDSD calculation gives a distance of $D$=4350 pc, with a one-sigma range from 4040 pc to 4860 pc.  Bailer-Jones et al. (2021) return a similar result ($D$=4056$\pm$440 pc) for assuming that KT Eri follows a spatial distribution like in their detailed galactic model.

However, this distance is greatly larger than the prior expectation of $L$=280 pc.  This likely points to the standard prior expectation for KT Eri as being wrong due to the nova not following the distribution of all the other disk novae.  But we cannot just adopt $L=(1000 pc)/\pi$, because this would be a circular bias.  Perhaps the most reasonable approach is to use the additional prior information that the peak absolute magnitudes of novae are in the range $-$7.0$\pm$1.4 (Schaefer 2018).  With extinction correction, this points to a distance of 2700 pc (with a one-sigma range of 1400--5100 pc).  For $L$=2700 pc, then the EDSD calculation gives $D$=5110$_{-430}^{+920}$ pc, and this is what we adopt.

\subsection{Absolute Magnitudes}

The absolute magnitude is important for determining the accretion rate in KT Eri.  Fortunately, we have a good distance from {\it Gaia} and the extinction has small uncertainty.  This gives $M_V$=$V$-13.80$\pm$0.25.

With $V_{peak}$=5.42$\pm$0.02, then $M_{V_{peak}}$ equals -8.38$\pm$0.25.  This is quite ordinary for nova eruptions, although a bit on the luminous side (Schaefer 2018).

For the application of finding the average $M_V$ over the eruption cycle, it is best to use the average V-magnitudes from 1895 to 2021 in quiescence.  The B-magnitudes are converted to V-magnitudes with the average post-eruption $\langle B-V \rangle$=0.23.  The V-band magnitudes were 14.77 in 1895, 14.30 in the 1910s, 13.96 in the 1920s, 14.08 in the 1930s, 14.52 in the 1940s, 14.85 in the early 1950s, 14.33 in the 1980s, 14.47 in the early 2000s, 15.09 in the last half of the 2000s before the eruption, and 15.06 in the decade after the end of the eruption.  With the vagaries of sampling, it is difficult to come up with a formal average or uncertainty.  For example, if we took a straight average, then this would heavily bias to faint magnitudes because the post-eruption decade has a very high density of magnitudes while it just happens to be systematically fainter than pre-eruption intervals.  A fairer method to find the century-long average magnitude would be to average the decadal averages.  Even this has a variety of reasonable calculations, for example by using the average or median, and whether or not we combine the separate values from the early and late decade starting in 2000.  Fortunately, all these methods return a middle or average value of $\langle V_q \rangle$=14.50 for the entire century-long light curve.  With the large variability in the intrinsic light curve and our sampling, the various ways to calculate the error bar are not convincing, yet such attempts return values close to $\pm$0.20.  This century-long average magnitude of 14.5$\pm$0.2 translates to an absolute visual magnitude of $\langle M_{V_q} \rangle$=0.7$\pm$0.3.

In Sections 3.6 and 3.9, we will be needing a range of absolute magnitude that is representative of the variations.  Above in this paper, we have presented an array or maximum and minimum values of $\langle V_q \rangle$ for various time intervals and for averaged or extreme magnitudes.  For purposes of calculating the recurrence time scale, the extremes are not appropriate, the decadal averages are hiding the times of high and low accretion, while we cannot select just the pre-eruption or post-eruption intervals.  What is needed is something like the magnitudes from 1895--2021 (outside eruption) for which only 10 per cent are fainter and only 10 per cent are brighter.  Again, we need to avoid doing straight counts in the light curve, as such would bias the range for those time intervals with a high density of magnitudes.  But monthly, yearly, or decadal groupings would average out much of the variability being measured.  With and without various weighting schemes, the dividing magnitude for the faintest 10 per cent is always near $V$=15.3, and the dividing magnitude for the brightest 10 per cent is roughly $V$=14.0. We are taking the typical range of variations in quiescence to be 14.0--15.3 mag.  The corresponding range of absolute magnitude is from $+$0.2 to $+$1.5 mag.

For purposes of calculating the accretion rate, we need the absolute magnitude of the disc alone, $M_{V_{disc}}$.  The correction from $M_{V_{disc}}$ to $M_{V_{disc}}$ was calculated with the SED fits for a blackbody plus $\alpha$-disk (Section 2.2).  The century-long average (with $\langle V_q \rangle$=14.5) is between the middle-SED case (with $\langle V_q \rangle$=15.00) and the brightest-SED case (with $\langle V_q \rangle$=13.88).  With an interpolation from our SED fits, the correction from the total magnitude to the disc-only magnitude is 0.25 mag.  This correction is taking out the light from the companion star.  If we only count the disc light in the century-long average case, the magnitude would be $V$=14.75. The corresponding $\langle M_{V_{disc}} \rangle$ is then +0.95 with an error bar near $\pm$0.3 mag.

The range of $M_{V_{disc}}$ is of relevance for seeing the effects of the variability of $\dot{M}$ for KT Eri.  From our SED fits (see Section 2.2),  the magnitude corrections to get the disc light alone are roughly +0.15 and +0.5 mag respectively. The typical variations in $M_{V_{disc}}$ is from +0.35 to +2.0 mag.

The magnitudes and absolute magnitudes are tabulated in Table 7.  This allows us to keep the values for the three states in order.  Further, this table illustrates the line of analysis, where the observed $V$-magnitudes are converted to absolute magnitudes, then the absolute magnitudes for the disc-light alone are separated out, then converted to an accretion rate and finally a recurrence time scale.

KT Eri in quiescence is at the top end of the $M_{V_q}$ distribution for novae.  Patterson et al. (2021) shows that the historical nova are from +3.0 to +6.0, with recurrent novae having $M_{V_q}$ around +1.0.  Schaefer (2018) found that 88 per cent of the CNe in the gold sample have $M_{V_q}$ less luminous than +3.0.  This result is pointing to KT Eri as a strong RN candidate.

\begin{table}
	\centering
	\caption{Calculation of $\tau_{rec}$ from observed magnitudes}
	\begin{tabular}{lrrr} 
		\hline
		 & Faintest & Average  &  Brightest   \\
		\hline
$V_q$ (mag)	&	15.3	&	14.5$\pm$0.2	&	14.0	\\
$M_{V_q}$ (mag)	&	$+$1.5	&	$+$0.7$\pm$0.3	&	$+$0.2	\\
$M_{V_{disc}}$ (mag)	&	$+$2.0	&	$+$0.95$\pm$0.3	&	$+$0.35	\\
$\dot{M}$ ($10^{-7}$ M$_{\odot}$/year)	&	0.83	&	3.5$^{+1.8}_{-1.2}$	&	8.2	\\
$\log$[$\dot{M}$] 	&	-7.08	&	$-$6.46$\pm$0.19	&	$-$6.09	\\
$\tau_{rec}$ (years)	&	50	&	12$\pm$7	&	5	\\
		\hline
	\end{tabular}

\end{table}

\subsection{The Accretion Rate}

The optical light from KT Eri is dominated by the accretion disc.  The luminosity of the accretion disc is proportional to the mass accretion rate.  Unfortunately, it is effectively impossible to directly measure the accretion luminosity because much of the flux comes out in the invisible ultraviolet.  Instead, the traditional solution is to use the $M_{V_{disc}}$ as a good proxy for $\dot{M}$.  Fortunately, the physics of cataclysmic variable accretion discs is well known as for the how to translated the $M_{V_{disc}}$ into $\dot{M}$ (e.g., Smak 1989; Frank et al. 2002; Dubus, Otulakowska-Hypka, \& Lasota 2018).  None of these references provides a conversion from $M_{V_{disc}}$ to $\dot{M}$ for the specific case of KT Eri, therefore, we have performed the identical calculation.  This is simply an integral across annuli in the standard $\alpha$-disc, each annulus having blackbody emission at a temperature falling off from the centre of the disc.  The standard equations appear many places (e.g., equations 5.43 and 5.45 of Frank et al. 2002).  This integral was performed numerically, resulting in a $F_{\nu}$ value in Janskys for 5500 \AA ~at a distance of 10 parsecs, with this then being converted to $M_{V_{disc}}$.  The most critical inputs are $\dot{M}$, M$_{WD}$, and the inclination.  Other inputs are the period, the companion mass, and the fraction of the white dwarf's Roche lobe filled by the accretion disc, with these three parameters used only to get a radius for the outer edge of the accretion disc, while the $M_{V_{disc}}$ has only negligible dependency on these three parameters for the situation of KT Eri.  As an ordinary check on our calculations, we are able to reproduce the results displayed in Smak (1989).  Our procedure is to use the best system parameters (from Table 3), introduce variations within the stated uncertainty range as quoted, and vary the $\dot{M}$ until the target $M_{V_{disc}}$ is reached.  Our derived accretion rates are given in Table 7.

The most important conversion is from the century-long average $\langle M_{V_{disc}} \rangle$=$+$0.95$\pm$0.30 to the $\langle \dot{M} \rangle$ as averaged over the entire eruption cycle.  With this, the average accretion rate over the eruption cycle is 3.5$^{+1.8}_{-1.2}$$\times$10$^{-7}$ M$_{\odot}$/year.  With frequently used logarithmic values, the rate is -6.46$\pm$0.18.  But the real uncertainty will be larger, due to the various uncertainties in the model input for translating $M_V$ into $\dot{M}$.  These errors can be estimated by looking at the variations in $\log$$\dot{M}$ for each input parameter.  The largest variations come from the error bars in $\langle M_{V_{disc}} \rangle$, as already expressed with the $\pm$0.18.  Varying the inclination over its allowed range of 47\degr to 57\degr, resulting in deviations of $\log$$\dot{M}$ by up to 0.07.  Variations in $\log$$\dot{M}$ are smaller than $\pm$0.02 for changes in $M_{WD}$, mass of the companion star, and the fraction of the Roche surface filled by the outer edge of the disc.  Added in quadrature, the error bars are $\pm$0.19. The final value for the eruption-cycle-averaged $\log$[$\langle \dot{M} \rangle$] is -6.46$\pm$0.19.  This is the relevant accretion rate for calculating the nova recurrence time-scale.

It is also useful to calculate the accretion rate for the extremes for the typical range in $M_{V_{disc}}$ from +0.35 to +2.0 mag.  For these two extremes, the accretion rate varies from 0.83$\times$10$^{-7}$ to 8.2$\times$10$^{-7}$ M$_{\odot}$/year.  The range of values of $\log$$\dot{M}$ is from -7.08 to -6.09 (with units of solar masses per year).

These $\dot{M}$ values are high, very high.  These accretion rates are higher than that of any other known nova.  The upper part of the derived $\dot{M}$ range is nominally within the zone of the `Nomoto plot' (Nomoto 1982) for which there is stable and continuing hydrogen burning on the surface of the white dwarf (see Fig. 8).  (The hatched zone is not from the Shen \& Bildsten full-physics calculation, but rather they took the analytic calculations from Nomoto et al. 2007.)  And for accretion rates $\gtrsim$6$\times$10$^{-7}$ M$_{\odot}$/year, the expectation is that the outer parts of the white dwarf will expand to red giant size, presumably creating a common envelope situation.  A possible explanation for this high mass transfer rate, may be what is called: `thermal time-scale mass transfer' (Ivanova \& Taam 2004). This can occur when the envelope of the donor becomes even just slightly out of thermal equilibrium, causing the envelope to expand. A sub-giant donor, just coming off the main-sequence, will have such a bloated envelope on its way to becoming a red giant. Since the accretion rate is highly sensitive to the radius of the donor (e.g., Ritter 1988), a bloating envelope will increase the Roche lobe overflow, and thus, enhance the rate of mass transfer, possibly explaining the high accretion rate we deduce here for KT Eri. Another possibility for a temporarily, thermally-induced, enhanced mass transfer, can be the irradiation of the donor from a previous nova eruption, causing the donor's envelope to slightly expand from the heat blast, and then slowly relax (Hillman et al., 2020, Hillman, 2021).  The operative scenario for systems like KT Eri is largely unknown, and there is no detailed physics for the accretion.  Unfortunately, for application to KT Eri, these theory predictions assume a steady accretion rate, while KT Eri has a rate that varies by an order of magnitude on most time-scales.  With the physical processes being non-linear, the theory cannot reach any conclusion for KT Eri.  We are not aware of any published models or precedents for how a star in this zone should appear or behave, much less one that is inside the zone only part of the time.  For KT Eri, some evidences give information on the accretion.  First, we see no sign of a heavy stellar wind.  Second, KT Eri is a  long-lasting supersoft X-ray source (Sun et al. 2020), which could point to sustained nuclear burning on the white dwarf.  Third, any episodic excursions to hydrogen burning cannot burn most of the accumulated hydrogen, because there was a normal nova eruption in 2009 during which strong hydrogen lines appear, while the helium lines are not enhanced (Munari et al. 2014).  

After the recurrent-nova-nature of KT Eri is established (see below), the front line for study of this system turns to the nature of the accretion, with this hopefully providing some sort of a ground-truth for systems with such high accretion.  With an eye towards the future of the KT Eri system, its evolution depends sensitively on the original main sequence mass of the companion star, with the end result possibly being a double white dwarf binary, an accretion induced collapse of the white dwarf, a Type {\rm I}a supernova, or a common envelope with a merger (see Fig. 4 of Ivanova \& Taam 2004).  Regardless of the uncertainties in the nature of the `stable hydrogen burning' regime, KT Eri has an accretion rate that is very very high.

\begin{figure}
	\includegraphics[width=1.01\columnwidth]{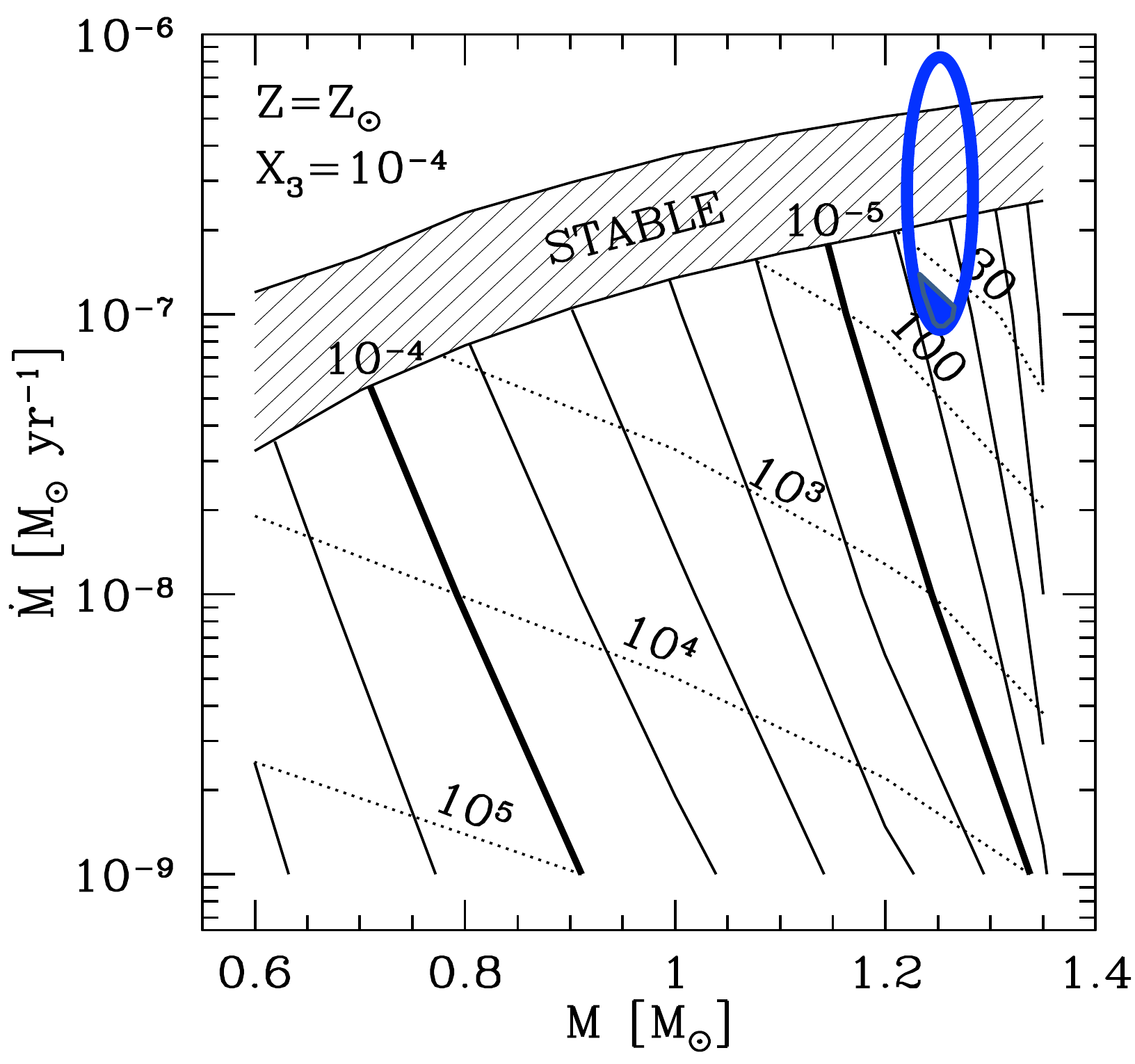}
    \caption{The `Nomoto plot' showing $\tau_{rec}$ as a function of $M_{WD}$ and $\dot{M}$.  The blue ellipse in the upper right shows the possible positions for KT Eri, with the vertical spread due to the range of intrinsic changes of $\dot{M}$ and the horizontal width due to the uncertainty in the white dwarf mass.  As KT Eri changes its accretion rate, the instantaneous position on this graph moves up and down within the ellipse.  Curves of constant $\tau_{rec}$ are shown as dotted lines running diagonally, with the 30-year-recurrence line in the upper right.  Just below this is the $\tau_{rec}$=100 year line that provides the threshold for the definition of a recurrent nova.  KT Eri is above this threshold in all cases, hence KT Eri is a recurrent nova.  While the nova moves from top to bottom within the ellipse, the operative level to trigger the eruption is at an unknown position that must be somewhere inside the ellipse.  From archival plates, there were no prior undiscovered nova eruptions from 1927--1954 or after 1969, within the blue ellipse of possibilities, $\tau_{rec}$ can only be from 40 to 50 years (as shown by the blue region cutting across the the bottom of the blue ellipse).  The thick and thin solid diagonal black lines show the trigger mass, M$_{trigger}$ in units of solar masses.  The hatched region across the top nominally has steady hydrogen burning on the white dwarf.  This Nomoto plot has the base diagram copied from Figure 7 of Shen \& Bildsten (2009), with solar abundances and the $^3$He mass fraction ($X_3$) equal to $10^{-4}$.}
\end{figure}

\subsection{White Dwarf Mass}

The mass of the white dwarf, $M_{WD}$, is critical for many of the central questions for KT Eri.  We have three methods to estimate $M_{WD}$:

The radial velocity curve has information on $M_{WD}$.  From Section 2.3, $K$=58.4$\pm$7.0 km s$^{-1}$ and P=2.615 days, for a mass function of 0.054$^{+0.022}_{-0.017}$ M$_{\odot}$.  For the nominal expected values of $M_{comp}$=1.0 M$_{\odot}$ and an inclination of 58$\degr$, the white dwarf is much more massive than the Chandrasekhar limit at 2.36 M$_{\odot}$.  If the input are individually push to their acceptable limits{\footnote{$K$ to 65.4 km s$^{-1}$, inclination to 51$\degr$, and $M_{comp}$ to 0.8 M$_{\odot}$}}, then the Chandraeskhar limit is still never approached.  If all three inputs are simultaneously pushed to their limits, then the white dwarf mass can be 0.98 M$_{\odot}$, which is certainly too low.  In all, the radial velocity curve does not provide any useful lower limit on $M_{WD}$.  Still, with the distributions for the inputs, a value just below the Chandrasekhar limit is the most likely.

I. Hachisu (University of Tokyo) and M. Kato (Keio University) have a long series of papers reporting their derivation of the white dwarf masses for many classical and recurrent novae (including Hachisu \& Kato 2016; 2018; 2019 and Hachisu, Kato, \& Schaefer 2003).  Their method is to make a detailed physics model of the entire system that addresses the multi-color light curves, including those in the optical, infrared, and X-ray.  Their methods yield results that closely match independent measures of $M_{WD}$.  One of their papers (Hachisu \& Kato 2018) specialized in deriving $M_{WD}$ for very-fast novae (just like KT Eri), including many recurrent novae.  On a query by email, Hachisu tells us that the KT Eri $M_{WD}$ had already been calculated.  Their best fitting value is 1.26 M$_{\odot}$, with an uncertainty of roughly 0.02 M$_{\odot}$.
	
Shara et al. (2018) presented a set of calculations for $M_{WD}$ and $\tau_{rec}$ for 82 CNe and 10 RNe in our own galaxy.  The observational basis for this was the light curves and tabulated properties in the catalog of Strope et al. (2010) and the comprehensive RN review of Schaefer (2010).  The only required input is the eruption amplitude ($V_q$-$V_{peak}$) and $t_2$.  For this, they used model light curve calculations, such as in Yaron et al. (2005) to produce model amplitudes and $t_2$ for a grid of models with varying $M_{WD}$ and $\dot{M}$.  This encapsulates the primary variables, with secondary conditions having small import.  (Still, extreme conditions are not covered. For example, the helium envelope of V445 Pup means that the model cannot readily apply.  Similarly, the extremely small and large eruption amplitudes of V888 Cen and CP Pup are outside the grid.)  The grid was constructed for models with varying $M_{WD}$ and $\dot{M}$ as input, but can then be made into a grid of model amplitude and $t_2$.  An observed nova is placed on to the grid.  The white dwarf mass and the recurrence time-scale for the model at that grid point (with interpolation as needed) is extracted.  The derived $M_{WD}$ values are in good agreement with the masses measured from radial velocity curves.

We have applied this particular procedure for KT Eri.  The input is $t_2$=6.6 days and that the amplitude varies from 8.7 to 9.6 mag, depending on when the quiescence is measured.  The white dwarf mass is determined to be 1.21--1.27 M$_{\odot}$, with this range representing the range of input amplitudes.  The uncertainty is $\pm$0.05 M$_{\odot}$.

A substantial potential problem arises because the accretion rate in KT Eri is extremely variable throughout quiescence, with this leading to possibly large and non-linear effects.  The theory models used for the KT Eri estimates all assumed a constant accretion rate.  With a fast changing $\dot{M}$, the thermal structure and diffusion rates in the accreted envelope will likely change greatly from that of a constant-$\dot{M}$ case, and there is no guarantee that the resultant $t_2$ and amplitude will be anywhere near to that of the constant-$\dot{M}$ case.  This uncertainty is likely to be particularly large for estimating the recurrence time-scale, therefore, we are not quoting any $\tau_{rec}$ value.  Still, we expect that our estimate of $M_{WD}$ should lie somewhere between the values for the two extremes in amplitude.  For this third method, $M_{WD}$ lies between 1.21--1.27 M$_{\odot}$.

With judicious weighing of our three estimates, we adopt $M_{WD}$ as 1.25$\pm$0.03 M$_{\odot}$.

\subsection{Companion Star Mass}

The mass of the companion star is poorly constrained, both for KT Eri and for other similar novae.  Here are the various constraints:  {\bf (1)} For novae with periods pointing to evolved companions, the estimated masses, in solar masses, are 0.88 for U Sco, 1.05 for QZ Aur, 0.87 for BT Mon (Ritter \& Kolb 2003), and 1.05 for V3890 Sgr (Miko{\l}ajewska et al. 2021), all with mass ratios around 0.8.  For KT Eri, with a mass ratio of 0.8, the companion to be roughly 1.0 M$_{\odot}$.  {\bf (2)} The mass ratio of KT Eri cannot be higher than some limit like 1.0, or the accretion would runaway with a fast exponential increase, and the companion must have something less than 1.25 M$_{\odot}$.  {\bf (3)} The core mass of a subgiant star has to be roughly 0.6 M$_{\odot}$, and the companion can only be substantially larger.  {\bf (4)} The companion has a temperature of 6200$\degr$, which would point to the star being more massive than our Sun given its subgiant status, and providing that irradiation is not changing the effective temperature much.  {\bf (5)} Our radial velocity curve for KT Eri places only weak constraints on the subgiant's mass.  For the best values from Table 3 plus the 58$^{+6}_{-7}$ degree inclination (see Section 3.2), the companion mass is 0.70$\pm$0.14 M$_{\odot}$.  But it is easy to push to lower inclinations to allow companion star masses $>$1.25 M$_{\odot}$.  In all, the approximate mass is 1.0$\pm$0.2 M$_{\odot}$.

\subsection{Recurrence Time-Scale}

The recurrence time-scale between nova eruptions, $\tau_{rec}$, is the system property that determines whether KT Eri is a recurrent nova or not.  A cutoff of 100 years is a traditional, arbitrary, and useful definition for RNe as a way of cutting the continuum of $\tau_{rec}$ for novae.  Much of our data collection and analysis has turned into a quest to measure a reliable value for $\tau_{rec}$.  The easiest way to measure $\tau_{rec}$ is to find two eruptions and use the intervening time interval.  But such is not possible for most cataclysmic variables, because the probabilities of discovering a second eruption is almost always low.  If the vagaries of chance observations does not come up lucky, then $\tau_{rec}$ information must come from elsewise.  

For KT Eri, we use two sources of information.  First, we use our examination of archival sky photographs worldwide to constrain the possible times of prior eruptions.  Second, we use our $M_{WD}$ and $\dot{M}$ values to derive $\tau_{rec}$.

For first source of information on $\tau_{rec}$ is the archival plates, as discussed in Section 2.1.3.  KT Eri did not have any eruptions from 1928 to 1954, and any eruptions after 1969 are very unlikely. KT Eri can only have a $\tau_{rec}$$>$82 years or 40$<$$\tau_{rec}$$<$55 years.

Our second method to get $\tau_{rec}$ is to calculate it from the our $M_{WD}$ and $\dot{M}$ values.  The trigger mass ($M_{trigger}$) is the amount of material that must be accreted onto the white dwarf for which a thermonuclear runaway is triggered.  With some average accretion rate, 
\begin{equation}
\tau_{rec} = M_{trigger} / \langle \dot{M} \rangle.
\end{equation}
The runaway is triggered when the pressure at the base of the newly accreted layer of material reaches a critical value, 
\begin{equation}
P_{crit} = (G M_{WD} M_{trigger}) / (4 \pi R_{WD}^4).
\end{equation}
The white dwarf radius, $R_{WD}$ is a well known function of $M_{WD}$.  Shen \& Bildsten (2009) point to $P_{crit}$ values within a factor-of-3 of 1$\times$10$^{19}$ dyne cm$^{-2}$.  They find that the metallicity dependence is weak, scaling only as Z$^{-0.2}$.  Still $P_{crit}$ varies with $\dot{M}$ and $M_{WD}$, but not by a lot.  For $P_{crit}$ as a constant appropriate for the middle of the CN/RN boundary, then the CN/RN distinction applies closely for the KT Eri case.  $P_{crit}$ can be evaluated for all the ten known Galactic RNe, for which $\tau_{rec}$ values are roughly known.  (For some RNe, like for U Sco, the $\tau_{rec}$ is known to 10 per cent and is stable in time.)  With this, $P_{crit}$  is varied until the ten RNe have nearly correct recurrence time-scales.  This value is within $\sim$20 per cent of 0.7$\times$10$^{19}$ dyne cm$^{-2}$.  (This choice gives 8 years for U Sco, 50 years for T CrB, 17 years for RS Oph, and all below 100 years.)  This is a short path to calculating $\tau_{rec}$ from $M_{WD}$ and $\dot{M}$ for KT Eri.

With the century-averaged accretion rate of $\log$[$\langle \dot{M} \rangle$]=$-$6.46$\pm$0.19, a 1.25 M$_{\odot}$ white dwarf with a radius of 3540 kilometers, and the $P_{crit}$=0.7$\times$10$^{19}$ dyne cm$^{-2}$, $\tau_{rec}$ is 12$\pm$5 years.  Varying the white dwarf mass from 1.22 to 1.28 M$_{\odot}$, the recurrence time varies from 16 to 8 years. For the century-long average brightness level, the derived recurrence time scale is 12$\pm$7 years.  This assumes that the trigger threshold is steadily accreting at its average rate.  

A substantial worry for all these theory calculations is the effect of the highly-variable accretion rate on to the white dwarf.  Alternating episodes of high and low accretion might well return a substantially different trigger condition than for some middling-and-constant accretion.  It is easy to imagine that the runaway would start during high accretion rates with a different-from-average trigger threshold, and it is easy to imagine that the low accretion rate intervals would allow the accreted layer to relax to a condition farther from the trigger threshold.  Fortunately, the effects that we can imagine are all bounded by the maximum and minimum accretion rates.  That is, the effects of the variable accretion rate will never be more than that of a steady maximal accretion, and the effects will never be less than that of a steady minimal accretion.  While being somewhere between the maximum and the minimum, the best estimate can now only be for the average accretion rate. Even with all our imagined effects of the variable accretion, the $\tau_{rec}$ value must still be between the values derived for the high and low accretion states.  This is why we have been carefully tracking the range of $V_q$ through to $\langle \dot{M} \rangle$.  For $\log$[$\dot{M}$] of -6.09 and -7.08, the recurrence times is between 5 and 50 years.  Our best measure of $\tau_{rec}$ is 12 years, but a range of 5--50 years is within our uncertainties.

The above calculation with a constant $P_{crit}$ is easy to calculate, simple to understand, and closely approximates the underlying physics.  Further, the above calculations have the huge advantage that they match the existence of the ten known Galactic RNe, and their known $\tau_{rec}$ values.  Still, the straight forward calculation of $\tau_{rec}$ has not covered all the physics effects.  To cover the exact physics, Shen \& Bildsten (2009) have detailed models of all the nuclear reaction rates, realistic convection models within the accreted layer, and much more.  Further, they present their results such that their model $\tau_{rec}$ can be evaluated for the $M_{WD}$ and $\dot{M}$ case of KT Eri.  These results are shown in their fig. 7, which is reproduced in our Fig. 8.  On top of this plot, a blue ellipse delineate the allowed location of KT Eri based on our measured $M_{WD}$ and $\dot{M}$.  The $\tau_{rec}$ values are displayed as the diagonal dotted lines in the  Shen \& Bildsten plot.  This graph provides a confirmation of our simpler calculation.  In particular, the best-fitting position has a recurrence time of close to 12 years, while the range of allowed $\tau_{rec}$ is the same.  This is all just a confirmation of our above simpler calculation.  In the end, for the theory calculation from our measured $M_{WD}$ and $\dot{M}$, the calculated $\tau_{rec}$ of KT Eri is 12 years, with an extreme possible range of 5--50 years.  So, by this alone, KT Eri is an RN.

But we still have to combine our two sources of information on $\tau_{rec}$.  The best estimate value of 12 years violates the limits from the archival record. The real value must come from the long-end of the allowed range.  The $>$82 year range allowed by the archival data is eliminated by the theory calculation.  In Fig. 8, the ellipse should be cutoff for recurrence times of $<$40 years and $>$55 years, as required by the archival data.  We are left with only 40$<$$\tau_{rec}$$<$50 years. Our final answer is that we infer that KT Eri has a recurrence time-scale of 40--50 years.

\section{Is KT Eri a Recurrent Nova?}

The recognition of RNe is important for a wide range of applications concerning the evolution of novae.  Only ten Galactic RNe are known, and each system is individually of high importance.  Now we can ask whether KT Eri is the eleventh Galactic recurrent nova?

The prior operational definition of RNe was something like `a binary with a white dwarf that has two or more observed thermonuclear nova eruptions'.  All classical novae recur, even if on time-scales up to several million years (Yaron et al. 2005),  some sort of a mandated threshold for $\tau_{rec}$ is needed.  The round number of one century is appropriate and traditional.  A better formal definition of RNe is that they are a `binary with a white dwarf that undergoes thermonuclear nova eruptions with $\tau_{rec}$$<$100 years'.

Previously, the only way to prove that $\tau_{rec}$$<$100 years has been to spot two or more nova events from the same system.  But this is inefficient, as the discovery rate for {\it second} eruptions on a  previously know system is amazingly low (Schaefer 2010).  Many nova systems with only one observed eruption are actually RNe for which all of the other eruptions in the century have been missed (Pagnotta \& Schaefer 2014).



\begin{table*}
	\centering
	\caption{Proxy Criteria for Recognizing RN candidates}
	\begin{tabular}{llll} 
		\hline
		\# & Criterion & KT Eri  &  RN?   \\
		\hline
1	&	Nova Amplitude $<$ $14.5-4.5\log(t_3)$	&	9.2 $<$ 9.5	&	Yes	\\
2	&	$P$$>$0.6 days	&	$P$=2.6 days	&	Yes	\\
3	&	$J-H$$>$0.7 \& $H-K$$>$0.1	&	$J-H$=0.11 \& $H-K$=0.07	&	No - but OK	\\
4	&	H$\alpha$ FWHM > 2000 km s$^{-1}$	&	3300 km s$^{-1}$	&	Yes	\\
5	&	High-Excitation lines (He {\rm II} and/or Fe {\it X})	&	He {\rm II}	&	Yes	\\
6	&	P class (plateau in light curve)	&	P or PP	&	Yes	\\
7	&	$M_{WD}$$>$1.20 M$_{\odot}$	&	M$_{WD}$=1.25 M$_{\odot}$	&	Yes	\\
8	&	$M_{V_q}$$<$3	&	$M_{V_q}$=0.70	&	Yes	\\
		\hline
	\end{tabular}

\end{table*}

\subsection{The Eight Proxy Criteria}

Over the past many decades, many workers have pointed to a wide variety of nova properties that were claimed to indicate that the system was actually an RN.  Pagnotta \& Schaefer (2014) have systematized the previous six RN-criteria, and applied it to all available nova systems.  These six criteria are itemized in Table 8.  Criteria 4, 5, 6, and the speed class ($t_3$) are just proxies for recognizing a high mass white dwarf.  Criteria 2, 3, and the eruption  amplitude are just proxies for recognizing a high accretion rate.

For KT Eri, all six criteria are evaluated, with the results shown in Table 8.  KT Eri easily passes all but one of the criteria.  (Criteria 3 is just a proxy for having an evolved companion which forces the large $P$, and hence this criterion is satisfied more fundamentally by passing Criterion 2.)  KT Eri is a very good RN candidate. 

We propose two new proxy criteria (see Table 8), where RNe should have $M_{WD}$>1.20 M$_{\odot}$ and RNe should have $M_{V_q}$$<$3.0 mag.  Both of these criteria are individually satisfied by all known RNe in our galaxy, and are individually satisfied by 5 known CNe with $M_{WD}$>1.27 M$_{\odot}$ and $M_{V_q}$ more luminous than 2.5 mag.

\subsection{The Recurrence Time -Scale for KT Eri}

A recurrent nova is defined as a system with nova events that has $\tau_{rec}$$<$100 years.  This paper is putting forth a new method for recognizing RNe, and that is to use the observed $\dot{M}$ and $M_{WD}$ to calculate $\tau_{rec}$.

This new method has the strength that it can be applied to most nova systems.  And there is no need to wait for a century and hope that the next eruption is not missed.  This new method has the strength that the empirical calibration of $P_{crit}$ with the known Galactic RNe matches up with the deep theory results from the nuclear reaction rate models.  We know of no way to impeach these results outside of the usual propagation of uncertainties on the input.

The long train of observations and analysis in this paper has all been leading up to applying this method for KT Eri.  We found that $M_{WD}$=1.25$\pm$0.03 M$_{\odot}$ and that the century-long average $\dot{M}$ is 3.5$\times$10$^{-7}$ M$_{\odot}$/year.  This results in a best estimate of $\tau_{rec}$=12$\pm$7 years, although with a possible range from 5--50 years.  This calculation is constrained by the archival data which demonstrates that $\tau_{rec}$ cannot be faster than 40 years and cannot be 55--82 years.  With these, the primary result of our paper is that the recurrence time-scale of KT Eri is 40--50 years.  KT Eri is certainly a recurrent nova, even though only one of its eruptions has been discovered.

\section{The Nature of KT Eri}

\begin{figure}
	\includegraphics[width=1.01\columnwidth]{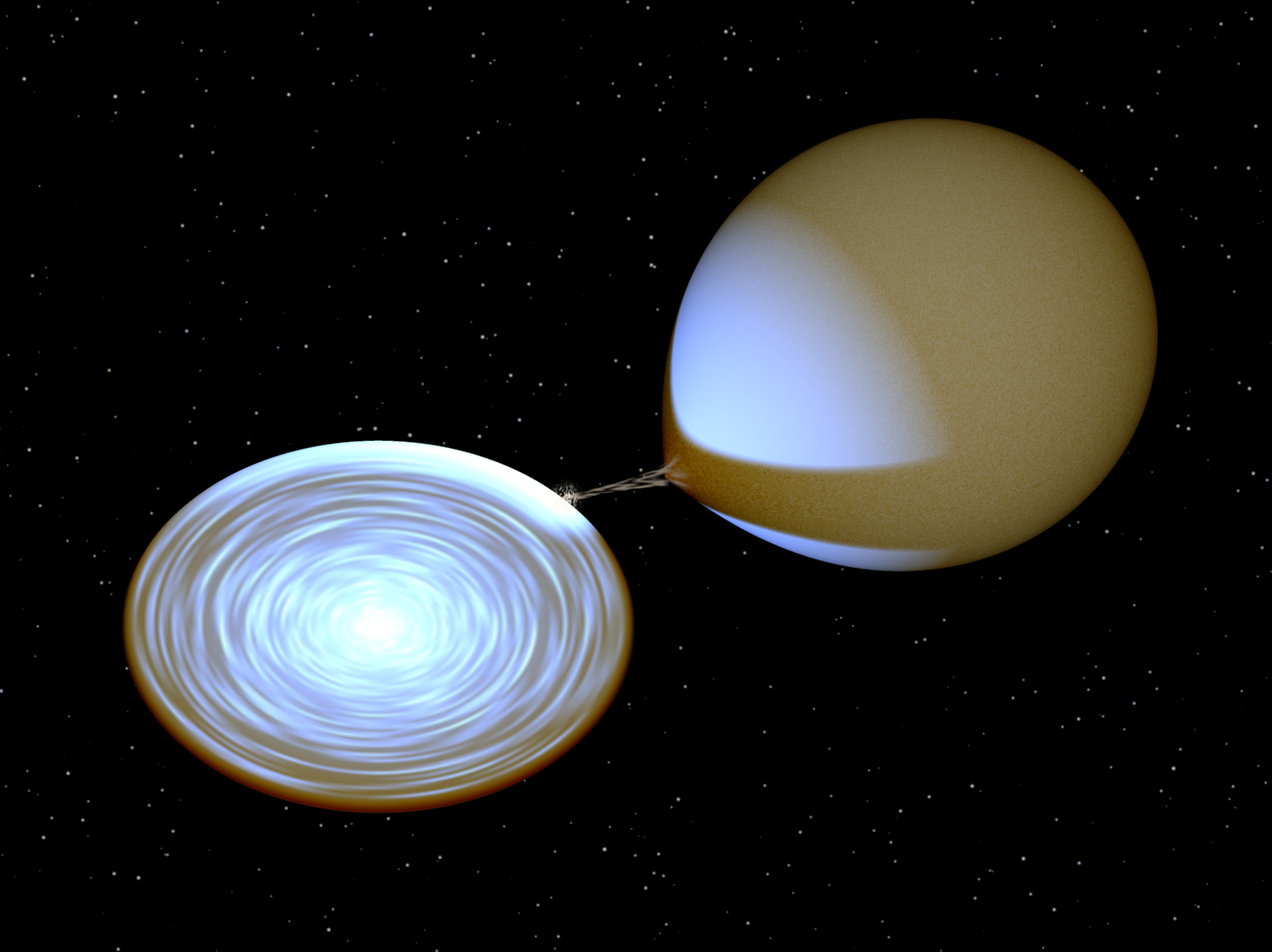}
    \caption{Close-up view of the KT Eri binary.  This picture is constructed with the correct sizes and distances for the stars, the positioning of the hotspot and accretion disc, and the viewing angle.  To set the scale, the companion star has a radius of 3.7 R$_{\odot}$.  The subgiant companion star has a temperature of 6200 K, while the irradiated polar regions appear brighter and hotter.  With the long orbital period, the accretion disc is very large, and the outer regions are relatively cool.  As the binary revolves, the extra light from the upper irradiated region will change in aspect, making for the brightness modulation on the orbital period.  In its average state, the disc outshines the companion star by 2:1.  The extreme variability of KT Eri likely results in a mottled disc.}
\end{figure}

An artistic visual picture of KT Eri is in Fig. 9, a realistic picture created by R. Hynes using his \texttt{BinSim} program{\footnote{http://www.phys.lsu.edu/~rih/binsim/}}.  Let us also give a verbal picture of the nature of KT Eri.  The binary has a period of 2.6 days, with the star centres separated by 10.4 R$_{\odot}$, and is viewed from a middle inclination.  The companion star has evolved off the main sequence and is now a subgiant star near 3.7 R$_{\odot}$ in radius with a temperature of 6200 K.  The companion star is losing mass through its inner Roche lobe at a very very high rate (3.5$\times$10$^{-7}$ M$_{\odot}$/year).  This accretion rate changes up and down by up to 10$\times$ on all time-scales from minutes to decades.  When the accretion rate is higher than average, KT Eri enters the `steady hydrogen burning' regime, although the understanding of this state is poor due to the rapid variability.  The accretion flow is through the usual disc, with no evidence for any magnetic field on the white dwarf.  The white dwarf is just below the Chandrasekhar mass, at 1.25$\pm$0.03 M$_{\odot}$.  The nova event of 2009 had a very high ejection velocity, a fast decline time, a `PP' plateau class light curve, and high excitation lines, all because the white dwarf is very massive.  The combination of a very high accretion rate plus a very high white dwarf mass requires that the nova recurrence time-scale must be 40--50 years, and KT Eri is a recurrent nova.

KT Eri is still `displaying' mysteries.  A detailed theory model is needed for how this system can have such a high accretion rate.  A theoretical understanding is needed for the large changes in the amplitude of the orbital modulations.   The extreme variability in this one system needs an explanation.  Detailed physics calculations are needed for what a system looks like in the `stable hydrogen burning' regime when the variability is so extreme.

\section{Acknowledgments}

We thank I. Hachisu and M. Kato for forwarding their calculation of the white dwarf mass of KT Eri.  We thank M. Shara for helpful discussions on white dwarf masses.  Rob Hynes provided the realistic \texttt{BinSim} image of the KT Eri system.  Research on novae at Stony Brook University has been supported by NSF grant AST1611443. Access to the SMARTS observing facilities has been made possible in part by research support from Stony Brook University.  The material is based upon work supported by NASA under award number 80GSFC21M0002.  This research has made use of the APASS database, located at the AAVSO web site; funding for APASS has been provided by the Robert Martin Ayers Sciences Fund.  This research made use of {\sc Lightkurve}, a Python package for Kepler and TESS data analysis (Lightkurve Collaboration, 2018).  The radial velocity analysis used software written in {\sc IDL}.  This work has made use of data from the European Space Agency (ESA) mission {\it Gaia}, processed by the {\it Gaia} Data Processing and Analysis Consortium (DPAC), with funding for the DPAC provided by national institutions, in particular the institutions participating in the {\it Gaia} Multilateral Agreement.

\section{Data Availability}

All of the photometry data used in this paper are explicitly given in Table 2, and is publicly available from the references and links in Table 1.


{}

\bsp	
\label{lastpage}

\begin{thebibliography}{99}

\bibitem[\protect\citeauthoryear{Bailer-Jones et al.}{2021}] {Bailer-Jones et al. 2021} 
Bailer-Jones C. A. L., Rybizki J., Fouesneau M., Demleitner M., Andrae R., 2021, AJ, 161, 147
\bibitem[\protect\citeauthoryear{Beardmore et al.}{2010}] {Beardmore et al. 2010} 
Beardmore A. P. et al., 2010, ATel, 2423
\bibitem[\protect\citeauthoryear{Bellm et al.}{2019}] {Bellm et al. 2019} 
Bellm E. C. et al., 2019, PASP, 131, 18002
\bibitem[\protect\citeauthoryear{Bode}{2010}] {Bode 2010} 
Bode M. F., 2010, ATel, 2392
\bibitem[\protect\citeauthoryear{Buffington et al.}{2007}] {Buffington et al. 2007} 
Buffington A., Morrill J. S., Hick P. P., Howard R. A., Jackson B. V., Webb D. F., 2007, Proc. SPIE, 6689, 66890B-1
\bibitem[\protect\citeauthoryear{Drake et al.}{2009}] {Drake et al. 2009} 
Drake A. J. et al., 2009, ATel, 2331
\bibitem[\protect\citeauthoryear{Drake et al.}{2014}] {Drake et al. 2014} 
Drake A. J. et al., 2014, MNRAS, 441, 1186
\bibitem[\protect\citeauthoryear{Dubus et al.}{2018}] {Dubus et al. 2018} 
Dubus G., Otulakowska-Hypka M., Lasota J.-P., 2018, A\&A, 617, 26
\bibitem[\protect\citeauthoryear{Frank et al.}{2002}] {Frank et al. 2002} 
Frank J., King A., Raine D., 2002, Accretion Power in Astrophysics, Cambridge Univ. Press, Cambridge
\bibitem[\protect\citeauthoryear{Gaia}{2021}] {Gaia 2021} 
{\it Gaia} Collaboration, 2021, A\&A, 649, A1
\bibitem[\protect\citeauthoryear{Hachisu and Kato}{2016}] {Hachisu and Kato 2016} 
Hachisu I., Kato M., 2016, ApJ, 816, 26
\bibitem[\protect\citeauthoryear{Hachisu and Kato}{2018}] {Hachisu and Kato 2018} 
Hachisu I., Kato M., 2018, ApJS, 237, 4
\bibitem[\protect\citeauthoryear{Hachisu and Kato}{2019}] {Hachisu and Kato 2019} 
Hachisu I., Kato M., 2019, ApJS, 242, 18
\bibitem[\protect\citeauthoryear{Hachisu et al.}{2003}] {Hachisu et al. 2003} 
Hachisu I., Kato M., Schaefer B. E., 2003, ApJ, 584, 1008
\bibitem[\protect\citeauthoryear{Henden et al.}{2012}] {Henden et al. 2012} 
Henden A. A., Levine S. E., Terrell D., Smith T. C., Welch D., 2012, JAAVSO, 40, 430
\bibitem[\protect\citeauthoryear{Hick et al.}{2007}] {Hick et al. 2007} 
Hick P. P., Buffington A., Jackson B. V. 2007, Proc. SPIE, 6689, 66890C-1
\bibitem[\protect\citeauthoryear{Hillman et al.}{2014}] {Hillman et al. 2014} 
Hillman Y., Prialnik D., Kovetz A., Shara M. M., Neill J. D., 2014, MNRAS, 437, 1962
\bibitem[\protect\citeauthoryear{Hillman et al.}{2020}] {Hillman et al. 2020} 
Hillman Y., Shara M. M., Prialnik D., Kovetz A., 2020, Nature Astron., 4, 886
\bibitem[\protect\citeauthoryear{Hillman}{2021}] {Hillman 2021} 
Hillman Y., 2021, MNRAS, 505, 3260
\bibitem[\protect\citeauthoryear{Hounsell et al.}{2010}] {Hounsell et al. 2010} 
Hounsell R. et al., 2010, ApJ, 724, 480
\bibitem[\protect\citeauthoryear{Ilkiewicz et al.}{2013}] {Ilkiewicz et al. 2013} 
Ilkiewicz  K., Swierczynski E., Galan C., Cikala M., Tomov T., 2014, in Woudt P. A., Ribeiro V. A. R. M., eds, ASP Conf. Ser. Vol. 490, Stella Novae: Future and Past Decades.  Astron. Soc. Pac., San Francisco, p. 411
\bibitem[\protect\citeauthoryear{Ivanova and Taam}{2004}] {Ivanova and Taam 2004} 
Ivanova N., Taam R. E., 2004, ApJ, 601, 1058
\bibitem[\protect\citeauthoryear{Jenkins et al.}{2016}] {Jenkins et al. 2016} 
Jenkins J. M., Twicken J. D., McCauliff S., Campbell J., 2016, Proc. SPIE, 9913, 99133E
\bibitem[\protect\citeauthoryear{Jurdana-Sepic et al.}{2012}] {Jurdana-Sepic et al. 2012} 
Jurdana-\v{S}epi\'{c} R., Ribeiro,V.A.R.M., Darnley M.J., Munari U., Bode M.F., 2012, A\&A, 537, A34
\bibitem[\protect\citeauthoryear{Kato and Hachisu}{1994}] {Kato and Hachisu 1994} 
Kato M., Hachisu I., 1994, ApJ, 437, 802
\bibitem[\protect\citeauthoryear{Lightkurve Collaboration}{2018}] {Lightkurve Collaboration 2018} 
Lightkurve Collaboration, 2018, Astrophysics Source Code Library, record ascl:1812.013
\bibitem[\protect\citeauthoryear{Lindegren et al.}{2021}] {Lindegren et al. 2021} 
Lindegren, L. et al., 2021, A\&A, 649,  A2
\bibitem[\protect\citeauthoryear{Luri et al.}{2018}] {Luri et al. 2018} 
Luri X. et al., 2018, A\&A, 616, A9
\bibitem[\protect\citeauthoryear{Mikolajewska et al.}{2021}] {Mikolajewska et al. 2021}Miko{\l}ajewska J. et al., 2021, MNRAS, 2021, 504, 2122\bibitem[\protect\citeauthoryear{Munari and Dallaporta}{2014}] {Munari and Dallaporta 2014} 
Munari U., Dallaporta S., 2014, New Astronomy, 27, 25
\bibitem[\protect\citeauthoryear{Munari et al.}{2014}] {Munari et al. 2014} 
Munari U., Mason E., Valisa P., 2014, A\&A, 564, 76
\bibitem[\protect\citeauthoryear{Ness et al.}{2015}] {Ness et al. 2015} 
Ness J.U. et al., 2015, A\&A, 578, A39
\bibitem[\protect\citeauthoryear{Nomoto}{1982}] {Nomoto 1982} 
Nomoto K., 1982, ApJ, 253, 798
\bibitem[\protect\citeauthoryear{Nomoto et al.}{2007}] {Nomoto et al. 2007} 
Nomoto K., Saio H., Kato M., Hachisu I., 2007, ApJ, 663, 1269
\bibitem[\protect\citeauthoryear{O'Brien et al.}{2010}] {O'Brien et al. 2010} 
O'Brien T. J. et al., 2010, ATel, 2434
\bibitem[\protect\citeauthoryear{Pagnotta and Schaefer}{2014}] {Pagnotta and Schaefer 2014} 
Pagnotta A., Schaefer B. E., 2014, ApJ, 788, 164
\bibitem[\protect\citeauthoryear{Pagnotta et al.}{2009}] {Pagnotta et al. 2009} 
Pagnotta A., Schaefer B. E., Xiao L., Collazzi A. C., Kroll P.,  2009, AJ, 138, 1230
\bibitem[\protect\citeauthoryear{Patterson}{1984}] {Patterson 1984}
Patterson J., 1984, ApJS, 54, 443
\bibitem[\protect\citeauthoryear{Patterson et al.}{2021}] {Patterson et al. 2021} 
Patterson J. et al., 2021, ApJ, in press, see arXiv:2010.07812
\bibitem[\protect\citeauthoryear{Pei et al.}{2021}] {Pei et al. 2021} 
Pei S., Orio M., Ness J.U., Ospina N., 2021, MNRAS, 507, 2073
\bibitem[\protect\citeauthoryear{Pojmanski}{1997}] {Pojmanski 1997} 
Pojmanski G., 1997, Act Astron., 47, 467
\bibitem[\protect\citeauthoryear{Rau et al.}{2009}] {Ragan 2009} 
Ragan E. et al., 2009, ATel, 2327
\bibitem[\protect\citeauthoryear{Raj et al.}{2013}] {Raj et al. 2013} 
Raj A., Banerjee D. P. K., Ashok N. M., 2013, MNRAS, 433, 2657
\bibitem[\protect\citeauthoryear{Ragan}{2009}] {Rau et al. 2009} 
Rau A. et al., 2009, PASP, 121, 1395
\bibitem[\protect\citeauthoryear{Ribeiro et al.}{2013}] {Ribeiro et al. 2013} 
 Ribeiro V. A. R. M., Bode M. F., Darnley M. J., Barnsley R. M., Munari U., Harman D. J.,  2013, MNRAS, 433, 1991
\bibitem[\protect\citeauthoryear{Ritter}{1988}] {Ritter 1988} 
Ritter H., 1988, A\&A, 202, 93
\bibitem[\protect\citeauthoryear{Ritter and Kolb}{2003}] {Ritter and Kolb 2003}Ritter H., Kolb U. 2003, A\&A, 404, 301 (updated RKcat version 7.24, 2016)
\bibitem[\protect\citeauthoryear{Schaefer}{2010}] {Schaefer 2010} 
Schaefer B. E.,  2010, ApJS, 187, 275
\bibitem[\protect\citeauthoryear{Schaefer}{2018}] {Schaefer 2018} 
Schaefer B. E., 2018, MNRAS, 481, 3033
\bibitem[\protect\citeauthoryear{Schaefer}{2021a}] {Schaefer 2021a} 
Schaefer B. E., 2021a, RNAAS, 5, 148
\bibitem[\protect\citeauthoryear{Schaefer}{2021b}] {Schaefer 2021b} 
Schaefer B. E., 2021b, RNAAS, 5, 150
\bibitem[\protect\citeauthoryear{Schaefer and Desai}{1988}] {Schaefer and Desai 1988} 
Schaefer B. E., Desai U.,  1988, A\&A, 195, 123
\bibitem[\protect\citeauthoryear{Schaefer et al.}{2010}] {Schaefer et al. 2010} 
Schaefer B. E. et al., 2010, ApJ, 742, 113
\bibitem[\protect\citeauthoryear{Schaefer et al.}{2013}] {Schaefer et al. 2013} 
Schaefer B. E. et al., 2013, ApJ, 773, 55
\bibitem[\protect\citeauthoryear{Schaefer}{2022}] {Schaefer 2022}
Schaefer B. E., Pagnotta A., Zoppelt S., 2022, MNRAS, 512, 1924
\bibitem[\protect\citeauthoryear{Schlafly and Finkbeiner}{2011}] {Schlafly and Finkbeiner 2011} 
Schlafly E.F., Finkbeiner D.P.,  2011, ApJ 737, 103
\bibitem[\protect\citeauthoryear{Shara et al.}{2018}] {Shara et al. 2018} 
Shara M. M., Prialnik D., Hillman Y., Kovetz A., 2018, ApJ, 860, 110
\bibitem[\protect\citeauthoryear{Shen and Bildsten}{2009}] {Shen and Bildsten 2009} 
Shen K. J., Bildsten L., 2009, ApJ, 699, 324
\bibitem[\protect\citeauthoryear{Smak}{1989}] {Smak 1989} 
Smak J., 1989, Acta Astron, 39, 317
\bibitem[\protect\citeauthoryear{Strope, Schaefer, and Henden}{2010}] {Strope, Schaefer, and Henden 2010} Strope R. J., Schaefer B. E., Henden A. A., 2010, AJ, 140, 34
\bibitem[\protect\citeauthoryear{Sun et al}{2020}] {Sun et al. 2020} 
Sun B., Orio M., Dobrotka A., Luna G. J. M., Shugarov S., Zemko P., 2020, MNRAS, 499, 3006
\bibitem[\protect\citeauthoryear{Walter et al.}{2012}] {Walter et al. 2012} 
Walter F. M., Battisti A., Towers S. E., Bond H. E., Stringfellow G. S., 2012, PASP, 124, 1057
\bibitem[\protect\citeauthoryear{Yaron et al.}{2005}] {Yaron et al. 2005} 
Yaron O., Prialnik D., Shara M. M., Kovetz A., 2005, ApJ, 623, 398


\end{thebibliography}
\end{document}